\documentclass[aps, prd, reprint, floatfix, superscriptaddress, nofootinbib,
showpacs, preprintnumbers, showkeys, 10pt]{revtex4-1}%
\usepackage{amsmath} \usepackage{amssymb} \usepackage{amsfonts}
\usepackage{amsthm} \usepackage{mathtools} \usepackage{mathrsfs}
\usepackage[caption=false, singlelinecheck=false,
justification=raggedright]{subfig}%
\usepackage{xspace} \usepackage{txfonts} \usepackage{pifont}
\usepackage{multirow} \usepackage{array} \usepackage[normalem]{ulem}%
\usepackage{rotating} 
\usepackage[colorlinks, urlcolor=black, citecolor=black, link
color=black]{hyperref}%
\renewcommand{\Re}{\ensuremath{\text{Re}}}
\renewcommand{\Im}{\ensuremath{\text{Im}}}

\newcommand{\nubar}{\ensuremath{\overline{\nu}}}

\newcommand{\modulus}[1]{\left| #1 \right|}

\newcommand{\bracket}[3]{\langle #1 \lvert #2 \rvert #3 \rangle}

\newcommand*{\DOI}[2]{\href{http://dx.doi.org/\detokenize{#1}}{#2}}

\newcommand{\CheckMark}{\textrm{\ding{51}}}
\newcommand{\CrossMark}{\textrm{\ding{55}}}

\allowdisplaybreaks

\begin{document}

\title{Angular distribution as an effective probe of new physics\\ in
	semi-hadronic three-body meson decays}%

\author{C.~S.~Kim} \email[E-mail at: ]{cskim@yonsei.ac.kr}%
\affiliation{Department of Physics and IPAP, Yonsei University, Seoul 120-749,
Korea}%

\author{Seong Chan Park}\email[E-mail at: ]{sc.park@yonsei.ac.kr}
\affiliation{Department of Physics and IPAP, Yonsei University, Seoul 120-749,
	Korea}%

\author{Dibyakrupa Sahoo} \email[E-mail at: ]{sahoodibya@yonsei.ac.kr\\S.C.P.\ and D.S.\ are the corresponding authors.}%
\affiliation{Department of Physics and IPAP, Yonsei University, Seoul 120-749,
Korea}%

\date{\today}

\begin{abstract}
We analyze, in a fully model-independent manner, the effects of new physics on a
few semi-hadronic three-body meson decays of the type $P_i \to P_f f_1 f_2$,
where $P_i, P_f$ are well chosen pseudo-scalar mesons and $f_{1,2}$ denote
fermions out of which at least one gets detected in experiments. We find that
the angular distribution of events of these decays can probe many interesting
new physics, such as the nature of the intermediate particle that can cause
lepton-flavor violation, or presence of heavy sterile neutrino, or new
intermediate particles, or new interactions. We also provide angular asymmetries
which can quantify the effects of new physics in these decays. We illustrate the
effectiveness of our proposed methodology with a few well chosen decay modes
showing effects of certain new physics possibilities without any hadronic
uncertainties.
\end{abstract}



\preprint{LDU-18-004}

\maketitle

\section{Introduction}

New physics (NP), or physics beyond the standard model, involves various models
that extend the well verified standard model (SM) of particle physics by
introducing a number of new particles with novel properties and interactions.
Though various aspects of many of these particles and interactions are
constrained by existing experimental data, we are yet to detect any definitive
signature of new physics in our experiments. Nevertheless, recent experimental
studies in $B$ meson decays, such as $B \to K^{(*)} \ell^-\ell^+$
\cite{B2KorKstLL}, $B_s \to \phi \ell^-\ell^+$ \cite{Aaij:2015esa}, $B \to
D^{(*)}\ell\nu$ \cite{B2DorDstLN} and $B_c \to J/\psi \ell\nu$
\cite{Aaij:2017tyk} (where $\ell$ can be $e,\mu$ or $\tau$) have reported
anomalous observations raising the expectation of discovery of new physics with
more statistical significance. In this context, model-independent studies of
such semi-leptonic three-body meson decay processes become important as they can
identify generic signatures of new physics which can be probed experimentally.
In this paper, we have analyzed the effects of new physics, in a
model-independent manner, on the angular distribution of a general semi-hadronic
three-body meson decay of the type $P_i \to P_f f_1 f_2$, where $P_i$ and $P_f$
are the initial and final pseudo-scalar mesons respectively, and $f_{1,2}$
denote fermions (which may or may not be leptons but not quarks) out of which at
least one gets detected experimentally. Presence of new interactions, or new
particles such as fermionic dark matter (DM) particles or heavy sterile
neutrinos or long lived particles (LLP) would leave their signature in the
angular distribution and we show by example how new physics contribution can be
quantified from angular asymmetries. Our methodology can be used for detection
of new physics in experimental study of various three-body pseudo-scalar meson
decays at various collider experiments such as LHCb and Belle~II.

The structure of our paper is as follows. In
Sec.~\ref{sec:Lagrangian-and-amplitude} we discuss the most general Lagrangian
and amplitude which include all probable NP contributions to our process under
consideration. The relevant details of kinematics is then described in
Sec.~\ref{sec:kinematics}. This is followed by a discussion on the angular
distribution and the various angular asymmetries in
Sec.~\ref{sec:ang-dist-asymmetries}. In Sec.~\ref{sec:example} we present a few
well chosen examples illustrating the effects of new physics on the angular
distribution. In Sec.~\ref{sec:conclusion} we conclude by summarizing the
important aspects of our methodology and its possible experimental realization.

\section{Most general Lagrangian and Amplitude}\label{sec:Lagrangian-and-amplitude}

Following the model-independent analysis of the decay $B \to D \ell^-\ell^+$ as
given in Ref.~\cite{Kim:2016zbg} and generalizing it for our process $P_i \to
P_f f_1 f_2$ where $P_{i,f}$ can be $B, B_s, B_c, D, K, \pi$ etc.\ as
appropriate and $f_1 f_2$ can be $\ell^- \ell^+$, $\ell \bar{\ell'}$, $\ell
\nu_{\ell}$, $\ell \nu_S$, $\ell f^{DM}$, $\nu_{\ell}\nubar_{\ell}$,
$\nu_S\nubar_{\ell}$, $\nu_{\ell}\nubar_S$, $\nu_S \nubar_S$, $f^{DM}
\bar{f}^{DM}$, $f_1^{DM} f_2^{DM}$, $f_1^{LLP} f_2^{LLP}$ (with
$\ell,\ell'=e,\mu,\tau$ denoting leptons, $\nu_S$ being sterile neutrino,
$f^{DM}_{1,2}$ as fermionic dark matter and $f_{1,2}^{LLP}$ as long lived
fermions)\footnote{It is clear that we can not only analyze processes allowed in
	the SM but also those NP contributions from fermionic dark matter in the final
	state as well as including flavor violation. Our analysis as presented in this
	paper is fully model-independent and general in nature.}, we can write down the
effective Lagrangian facilitating the decay under consideration as follows,
\begin{eqnarray}
\mathcal{L}_{\textrm{eff}} &=& J_S \left( \bar{f}_1 f_2 \right)
+ J_P \left( \bar{f}_1~\gamma^5~f_2 \right) + \left(J_V\right)_{\alpha}
\left( \bar{f}_1~\gamma^{\alpha}~f_2 \right) \nonumber\\%
&& + \left(J_A\right)_{\alpha} \left( \bar{f}_1~\gamma^{\alpha}\gamma^5~f_2
\right) + \left(J_{T_1}\right)_{\alpha\beta} \left(
\bar{f}_1~\sigma^{\alpha\beta}~f_2 \right) \nonumber\\%
&& + \left(J_{T_2}\right)_{\alpha\beta} \left(
\bar{f}_1~\sigma^{\alpha\beta}\gamma^5~f_2 \right) + \text{h.c.},
\label{eq:Effective-Lagrangian}
\end{eqnarray}
where $J_S$, $J_P$, $\left(J_V\right)_{\alpha}$, $\left(J_A\right)_{\alpha}$,
$\left(J_{T_1}\right)_{\alpha\beta}$, $\left(J_{T_2}\right)_{\alpha\beta}$ are
the different hadronic currents which effectively describe the quark level
transitions from $P_i$ to $P_f$ meson. It should be noted that we have kept both
$\sigma^{\alpha\beta}$ and $\sigma^{\alpha\beta}\gamma^5$ terms. This is because
of the fact that the currents $\bar{f}_1 \, \sigma^{\alpha\beta} \, f_2$ and
$\bar{f}_1 \, \sigma^{\alpha\beta}\gamma^5 \, f_2$ describe two different
physics aspects namely the magnetic dipole and electric dipole contributions
respectively. In the SM, vector and axial-vector currents (mediated by photon,
$W^{\pm}$ and $Z^0$ bosons) and the scalar current (mediated by Higgs boson)
contribute. So every other term in Eq.~\eqref{eq:Effective-Lagrangian} except
the ones with $J_S$, $\left(J_V\right)_{\alpha}$ and $\left(J_A\right)_{\alpha}$
can appear in some specific NP model. Since, in this paper, we want to
concentrate on a fully model-independent analysis to get generic signatures of
new physics, we shall refrain from venturing into details of any specific NP
model, which nevertheless are also useful. It is important to note that $J_S$,
$\left(J_V\right)_{\alpha}$ and $\left(J_A\right)_{\alpha}$ can also get
modified due to NP contributions.

\begin{figure}[hbtp]
\centering%
\includegraphics[scale=1]{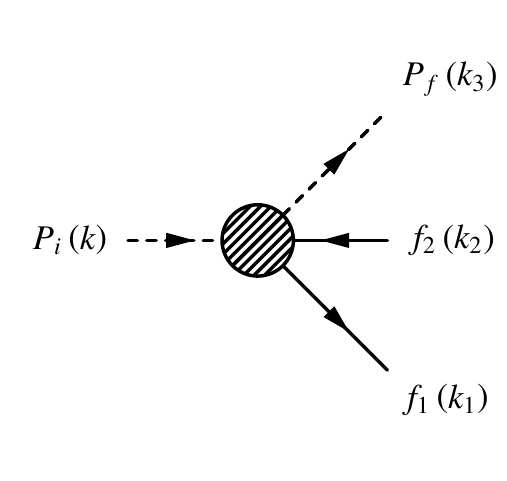} \caption{Feynman diagram
	for $P_i \to P_f f_1 f_2$ considering $f_1$ as a particle and $f_2$ as an
	anti-particle. Here the blob denotes the effective vertex and includes
	contributions from all the form factors defined in
	Eq.~\eqref{eq:form-factors}.}%
\label{fig:Feynman_diagram}
\end{figure}

In order to get the most general amplitude for our process under consideration,
we need to go from the effective quark-level description of
Eq.~\eqref{eq:Effective-Lagrangian} to the meson level description by defining
appropriate form factors. It is easy to write down the most general form of the
amplitude for the process $P_i \to P_f f_1 f_2$ depicted in
Fig.~\ref{fig:Feynman_diagram} as follows,
\begin{align}
\mathcal{M} \left( P_i \to P_f f_1 f_2 \right) &= F_S \left(
\bar{f}_1 f_2 \right) + F_P \left( \bar{f}_1~\gamma^5~f_2 \right)
\nonumber\\*%
&\quad + \left( F_V^+ p_{\alpha} + F_V^- q_{\alpha} \right) \left(
\bar{f}_1~\gamma^{\alpha}~f_2 \right) \nonumber\\* %
&\quad + \left( F_A^+ p_{\alpha} + F_A^- q_{\alpha} \right) \left(
\bar{f}_1~\gamma^{\alpha}~\gamma^5~f_2 \right) \nonumber\\* %
&\quad + F_{T_1}~p_{\alpha}~q_{\beta} \left( \bar{f}_1~\sigma^{\alpha\beta}~f_2
\right) \nonumber\\* %
&\quad + F_{T_2}~p_{\alpha}~q_{\beta} \left(
\bar{f}_1~\sigma^{\alpha\beta}~\gamma^5~f_2 \right), \label{eq:amplitude}
\end{align}
where $F_{S}$, $F_{P}$, $F_{V}^{\pm}$, $F_{A}^{\pm}$, $F_{T_1}$ and $F_{T_2}$
are the relevant form factors, and are defined as follows,
\begin{subequations}\label{eq:form-factors}
\begin{align}
\bracket{P_f}{J_S}{P_i} &= F_S,\\%
\bracket{P_f}{J_P}{P_i} &= F_P,\\%
\bracket{P_f}{\left(J_V\right)_{\alpha}}{P_i} &= F_V^+ p_{\alpha} + F_V^-
q_{\alpha},\\%
\bracket{P_f}{\left(J_A\right)_{\alpha}}{P_i} &= F_A^+ p_{\alpha} + F_A^-
q_{\alpha},\\%
\bracket{P_f}{\left(J_{T_1}\right)_{\alpha\beta}}{P_i} &=
F_{T_1}~p_{\alpha}~q_{\beta},\\%
\bracket{P_f}{\left(J_{T_2}\right)_{\alpha\beta}}{P_i} &=
F_{T_2}~p_{\alpha}~q_{\beta},
\end{align}
\end{subequations}
with $p \equiv k + k_3$ and $q \equiv k - k_3 = k_1 + k_2$, in which $k, k_1,
k_2, k_3$ are the 4-momenta of the $P_i, f_1, f_2 $ and $P_f$ respectively (see
Fig.~\ref{fig:Feynman_diagram}). All the form factors appearing in the amplitude
in Eq.~\eqref{eq:amplitude} and as defined in Eq.~\eqref{eq:form-factors} are,
in general, complex and contain all NP information. It should be noted that for
simplicity we have implicitly put all the relevant Cabibbo-Kobayashi-Maskawa
matrix elements as well as coupling constants and propagators inside the
definitions of these form factors. In the SM only $F_V^{\pm}$ and $F_A^{\pm}$
are present. Presence of NP can modify these as well as introduce other form
factors\footnote{It should be noted that the form factors, especially the ones
	describing semi-leptonic $B$ meson decays, can be obtained by using the heavy
	quark effective theory \cite{HQET}, the lattice QCD \cite{Lattice}, QCD
	light-cone sum rule \cite{Light-cone} or the covariant confined quark model
	\cite{CCQM} etc. In this paper we present a very general analysis which is
	applicable to a diverse set of meson decays. Hence we do not discuss any
	specifics of the form factors used in our analysis. Moreover, we shall show, by
	using certain examples and in a few specific cases, that one can also probe new
	physics without worrying about the details of the form factors. Nevertheless,
	when one concentrates on a specific decay mode, considering the form factors in
	detail is always useful.}. These various NP contributions would leave behind
their signatures in the angular distribution for which we need to specify the
kinematics in a chosen frame of reference.

\section{Decay Kinematics}\label{sec:kinematics}

\begin{figure}[hbtp]
\centering%
\includegraphics[scale=1]{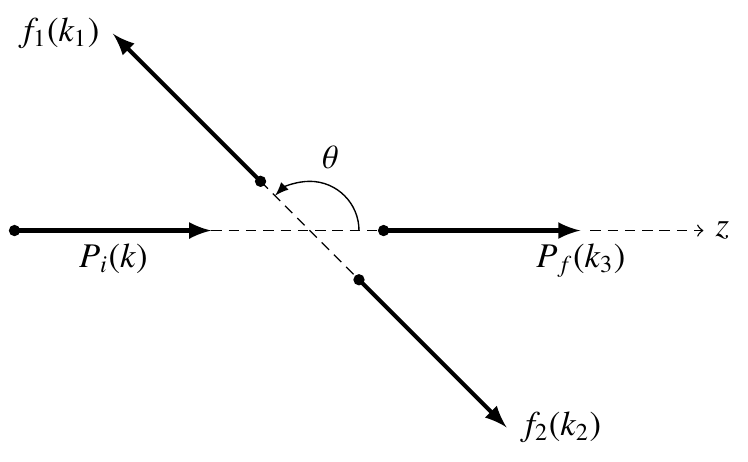}%
\caption{Decay of $P_i \to P_f f_1 f_2$ in the Gottfried-Jackson frame.} %
\label{fig:GJ-frame}
\end{figure}

We shall consider the decay $P_i \to P_f f_1 f_2$ in the Gottfried-Jackson
frame, especially the center-of-momentum frame of the $f_1,f_2$ system, which is
shown in Fig.~\ref{fig:GJ-frame}. In this frame the parent meson $P_i$ flies
along the positive $z$-direction with 4-momentum $k = \left(E, \mathbf{k}
\right) = \left(E,0,0,\modulus{\mathbf{k}}\right)$ and decays to the daughter
meson $P_f$ which also flies along the positive $z$-direction with 4-momentum
$k_3 = \left( E_3, \mathbf{k}_3 \right) = \left(E_3, 0, 0,
\modulus{\mathbf{k}_3}\right)$ and to $f_1$, $f_2$ which fly away back-to-back
with 4-momenta $k_1 = \left( E_1, \mathbf{k}_1 \right)$ and $k_2 = \left( E_2,
\mathbf{k}_2 \right)$ respectively, such that by conservation of 4-momentum we
get, $\mathbf{k}_1 + \mathbf{k}_2 = \mathbf{0}$, $\mathbf{k} = \mathbf{k}_3$,
and $E = E_1 + E_2 + E_3$. The fermion $f_1$ (which we assume can be observed
experimentally) flies out subtending an angle $\theta$ with respect to the
direction of flight of the $P_i$ meson, in this Gottfried-Jackson frame. The
three invariant mass-squares involved in the decay under consideration are
defined as follows,
\begin{subequations}\label{eq:stu}
\begin{align}
s &= (k_1 + k_2)^2 = (k - k_3)^2,\\%
t &= (k_1 + k_3)^2 = (k - k_2)^2,\\%
u &= (k_2 + k_3)^2 = (k - k_1)^2.
\end{align}
\end{subequations}
It is easy to show that $s + t + u = m_i^2 + m_f^2 + m_1^2 + m_2^2$, where
$m_i, m_f,m_1$ and $m_2$ denote the masses of particles $P_i,P_f,f_1$ and $f_2$
respectively. In the Gottfried-Jackson frame, the expressions for $t$ and $u$
are given by
\begin{subequations}\label{eq:tu}
\begin{align}
t &= a_t - b \cos\theta,\label{eq:t}\\%
u &= a_u + b \cos\theta,\label{eq:u}
\end{align}
\end{subequations}
where
\begin{subequations}\label{eq:ab}
\begin{align}
a_t &= m_1^2 + m_f^2 + \frac{1}{2s} \left( s + m_1^2 - m_2^2 \right) \left(
m_i^2 - m_f^2 -s \right),\label{eq:at}\\%
a_u &= m_2^2 + m_f^2 + \frac{1}{2s} \left( s - m_1^2 + m_2^2 \right) \left(
m_i^2 - m_f^2 -s \right),\label{eq:au}\\%
b &= \frac{1}{2s} \sqrt{\lambda\left( s, m_1^2, m_2^2 \right)~\lambda \left( s,
	m_i^2, m_f^2 \right)},\label{eq:b}
\end{align}
\end{subequations}
with the K\"{a}ll\'{e}n function $\lambda(x,y,z)$ defined as,
\begin{equation*}
\lambda\left( x,y,z \right) = x^2 + y^2 + z^2 - 2 \left( xy + yz + zx \right).
\end{equation*}
It is clear that $a_t$, $a_u$ and $b$ are functions of $s$ only. For the special
case of $m_1 = m_2 = m$ (say) we have $a_t = a_u = \tfrac{1}{2} \left(m_i^2 +
m_f^2 + 2m^2 -s\right)$ and $b = \tfrac{1}{2} \sqrt{\left(1-4m^2/s\right)
	\lambda\left(s,m_i^2,m_f^2\right)}$. It is important to note that we shall use
the angle $\theta$ in our angular distribution.

\section{Most general angular distribution and angular asymmetries}\label{sec:ang-dist-asymmetries}

Considering the amplitude as given in Eq.~\eqref{eq:amplitude}, the most general
angular distribution in the Gottfried-Jackson frame is given by,
\begin{equation}\label{eq:gen-angular-dist}
\frac{d^2\Gamma}{ds \, d\cos\theta} = \frac{b\sqrt{s} \left( C_0 + C_1
	\cos\theta + C_2 \cos^2\theta \right)}{128 \, \pi^3 \, m_i^2 \left(m_i^2 - m_f^2
	+ s \right)},
\end{equation}
where $C_0$, $C_1$ and $C_2$ are functions of $s$ and are given by,
\begin{subequations}\label{eq:gen-C012}
\begin{align}
C_0 &= 2 \Bigg(-\modulus{F_{T_1}}^2 \bigg(-\Sigma m_{12}^2 s^2 + 2 \Sigma
m_{12}^2 \left( \Sigma m^2 \right)_{if} s \nonumber\\%
& \hspace{1.5cm} + \left( \Delta m^2 \right)_{12}^2 s -\Delta a_{tu}^2 s - 2
\left( \Delta m^2 \right)_{12}^2 \left( \Sigma m^2 \right)_{if} \nonumber\\%
& \hspace{1.5cm} - \left( \Delta m^2 \right)_{if}^2 \Sigma m_{12}^2 + 2 \Delta
a_{tu} \left( \Delta m^2 \right)_{12} \left( \Delta m^2 \right)_{if} \bigg)
\nonumber\\%
& \quad - 2 \Im\left( F_V^+ F_{T_1}^* \right) \bigg( -\Sigma m_{12} s^2 + 2
\Sigma m_{12} \left( \Sigma m^2 \right)_{if} s \nonumber\\%
& \hspace{1.5cm} + \Delta m_{12} \left( \Delta m^2 \right)_{12} s - 2 \Delta
m_{12} \left( \Delta m^2 \right)_{12} \left( \Sigma m^2 \right)_{if}
\nonumber\\%
& \hspace{1.5cm} - \left( \Delta m^2 \right)_{if}^2 \Sigma m_{12} +\Delta a_{tu}
\Delta m_{12} \left( \Delta m^2 \right)_{if} \bigg) \nonumber\\%
& \quad + \modulus{F_{T_2}}^2 \bigg( \Delta m_{12}^2 s^2 - 2 \Delta m_{12}^2
\left( \Sigma m^2 \right)_{if} s - \left( \Delta m^2 \right)_{12}^2 s
\nonumber\\%
& \hspace{1.5cm} + \Delta a_{tu}^2 s + 2 \left( \Delta m^2 \right)_{12}^2 \left(
\Sigma m^2 \right)_{if} + \Delta m_{12}^2 \left( \Delta m^2 \right)_{if}^2
\nonumber\\%
& \hspace{1.5cm} - 2 \Delta a_{tu} \left( \Delta m^2 \right)_{12} \left( \Delta
m^2 \right)_{if} \bigg) \nonumber\\%
& \quad - 2 \Im\left( F_A^+ F_{T_2}^* \right) \bigg(\Delta m_{12} s^2 - 2 \Delta
m_{12} \left( \Sigma m^2 \right)_{if} s \nonumber\\%
& \hspace{1.5cm} - \left( \Delta m^2 \right)_{12} \Sigma m_{12} s + 2 \left(
\Delta m^2 \right)_{12} \Sigma m_{12} \left( \Sigma m^2 \right)_{if}
\nonumber\\%
& \hspace{1.5cm} - \Delta a_{tu} \left( \Delta m^2 \right)_{if} \Sigma m_{12} +
\Delta m_{12} \left( \Delta m^2 \right)_{if}^2 \bigg) \nonumber\\%
& \quad + \modulus{F_A^+}^2 \bigg( s^2 - 2\left( \Sigma m^2 \right)_{if} s -
\Sigma m_{12}^2 s \nonumber\\%
& \hspace{1.5cm} + 2 \Sigma m_{12}^2 \left( \Sigma m^2 \right)_{if} + \left(
\Delta m^2 \right)_{if}^2-\Delta a_{tu}^2 \bigg) \nonumber\\%
& \quad + \modulus{F_V^+}^2 \bigg( s^2 - 2 \left( \Sigma m^2 \right)_{if} s
-\Delta m_{12}^2 s \nonumber\\%
& \hspace{1.5cm} + 2 \Delta m_{12}^2 \left( \Sigma m^2 \right)_{if} + \left(
\Delta m^2 \right)_{if}^2 - \Delta a_{tu}^2 \bigg) \nonumber\\%
& \quad + \modulus{F_A^-}^2 \left(\Sigma m_{12}^2 s - \left( \Delta m^2
\right)_{12}^2 \right) \nonumber\\%
& \quad - 2 \Re\left( F_P F_A^{-*} \right) \left(\Sigma m_{12} s - \Delta m_{12}
\left( \Delta m^2 \right)_{12} \right) \nonumber\\%
& \quad - \modulus{F_V^-}^2 \left( \left( \Delta m^2 \right)_{12}^2-\Delta
m_{12}^2 s \right) \nonumber\\%
& \quad - 2\Re\left( F_S F_V^{-*} \right) \left(\left( \Delta m^2 \right)_{12}
\Sigma m_{12}-\Delta m_{12} s \right) \nonumber\\%
& \quad -\modulus{F_S}^2 \left( \Sigma m_{12}^2-s \right) -\modulus{F_P}^2
\left( \Delta m_{12}^2-s \right) \nonumber\\%
& \quad + 2 \Re\left( F_A^+ F_A^{-*} \right) \left( \left( \Delta m^2
\right)_{if} \Sigma m_{12}^2 - \Delta a_{tu} \left( \Delta m^2 \right)_{12}
\right) \nonumber\\%
& \quad - 2 \Re\left( F_P F_A^{+*} \right) \left( \left( \Delta m^2 \right)_{if}
\Sigma m_{12} - \Delta a_{tu} \Delta m_{12} \right) \nonumber\\%
& \quad - 2 \Re\left( F_S F_V^{+*} \right) \left( \Delta a_{tu} \Sigma m_{12} -
\Delta m_{12} \left( \Delta m^2 \right)_{if} \right) \nonumber\\%
& \quad + 2 \Re\left( F_V^+ F_V^{-*} \right) \left( \Delta m_{12}^2 \left(
\Delta m^2 \right)_{if} - \Delta a_{tu} \left( \Delta m^2 \right)_{12} \right)
\Bigg),\\%
C_1 &= 8 b \Bigg( \Delta m_{12} \left( \Im\left( F_V^- F_{T_1}^* \right)
s-\Re\left( F_P F_A^{+*} \right) \right) \nonumber\\%
& \hspace{1.5cm} + \Sigma m_{12} \Big(-\Im\left( F_A^- F_{T_2}^* \right) s +
\Re\left( F_S F_V^{+*} \right) \nonumber\\%
& \hspace{4cm} - \left( \Delta m^2 \right)_{if} \Im\left( F_A^+ F_{T_2}^*
\right) \Big) \nonumber\\%
& \hspace{1.5cm} + \Delta a_{tu} \left( \modulus{F_V^+}^2 + \modulus{F_A^+}^2
\right) \nonumber\\%
& \hspace{1.5cm} + \left( \Im\left( F_S F_{T_1}^* \right) + \Im\left( F_P
F_{T_2}^* \right) \right)s \nonumber\\%
& \hspace{1.5cm} + \left( \Delta m^2 \right)_{12} \left( \Re\left( F_V^+
F_V^{-*} \right) + \Re\left( F_A^+ F_A^{-*} \right) \right) \nonumber\\%
& \hspace{1.5cm} + \left( \Delta m^2 \right)_{if} \Delta m_{12} \Im\left( F_V^+
F_{T_1}^* \right) \Bigg),\\%
C_2 &= 8 b^2 \left( \left( \modulus{F_{T_2}}^2 + \modulus{F_{T_1}}^2 \right) s -
\modulus{F_V^+}^2 - \modulus{F_A^+}^2 \right),
\end{align}
\end{subequations}
with
\begin{subequations}
\begin{align}
\Delta a_{tu} &= a_t - a_u,\\%
\Delta m_{12} &= m_1 - m_2,\\%
\Delta m_{if} &= m_i - m_f,\\%
\Sigma m_{12} &= m_1 + m_2,\\%
\Sigma m_{if} &= m_i + m_f,\\%
\left(\Delta m^2\right)_{12} &= \Delta m_{12} \Sigma m_{12} = m_1^2 - m_2^2,\\%
\left(\Delta m^2\right)_{if} &= \Delta m_{if} \Sigma m_{if} = m_i^2 - m_f^2,\\%
\left(\Sigma m^2\right)_{if} &= m_i^2 + m_f^2.
\end{align}
\end{subequations}

In the limit $m_1=m_2$, which happens when $f_1 f_2 = \ell^-\ell^+, \nu\nubar,$
or $f^{DM} \bar{f}^{DM}$ etc., our expressions for the angular distribution
matches with the corresponding expression in Ref.~\cite{Kim:2016zbg}. It is
important to remember that in the SM we come across scalar, vector and axial
vector currents only. Therefore, in the SM, $F_P^{\text{SM}} =
F_{T_1}^{\text{SM}} = F_{T_2}^{\text{SM}} = 0$, which implies that,
\begin{subequations}\label{eq:SM-C012}
\begin{align}
C_0^{\text{SM}} =& 2 \Bigg( \modulus{\left(F_A^+\right)_{\text{SM}}}^2 \bigg(
s^2 - 2\left( \Sigma m^2 \right)_{if} s - \Sigma m_{12}^2 s \nonumber\\%
& \hspace{2cm} + 2 \Sigma m_{12}^2 \left( \Sigma m^2 \right)_{if} + \left(
\Delta m^2 \right)_{if}^2-\Delta a_{tu}^2 \bigg) \nonumber\\%
& \quad + \modulus{\left(F_V^+\right)_{\text{SM}}}^2 \bigg( s^2 - 2 \left(
\Sigma m^2 \right)_{if} s -\Delta m_{12}^2 s \nonumber\\%
& \hspace{2.25cm} + 2 \Delta m_{12}^2 \left( \Sigma m^2 \right)_{if} + \left(
\Delta m^2 \right)_{if}^2 - \Delta a_{tu}^2 \bigg) \nonumber\\%
& \quad + \modulus{\left(F_A^-\right)_{\text{SM}}}^2 \left(\Sigma m_{12}^2 s -
\left( \Delta m^2 \right)_{12}^2 \right) \nonumber\\%
& \quad - \modulus{\left(F_V^-\right)_{\text{SM}}}^2 \left( \left( \Delta m^2
\right)_{12}^2-\Delta m_{12}^2 s \right) \nonumber\\%
& \quad - \modulus{\left(F_S\right)_{\text{SM}}}^2 \left( \Sigma m_{12}^2 -s
\right) \nonumber\\%
& \quad + 2 \Re\left( \left(F_A^+\right)_{\text{SM}}
\left(F_A^-\right)_{\text{SM}}^* \right) \bigg( \left( \Delta m^2 \right)_{if}
\Sigma m_{12}^2 \nonumber\\*%
& \hspace{4cm} - \Delta a_{tu} \left( \Delta m^2 \right)_{12} \bigg)
\nonumber\\%
& \quad + 2 \Re\left( \left(F_V^+\right)_{\text{SM}}
\left(F_V^-\right)_{\text{SM}}^* \right) \bigg( \left( \Delta m^2 \right)_{if}
\Delta m_{12}^2 \nonumber\\%
& \hspace{4cm} - \Delta a_{tu} \left( \Delta m^2 \right)_{12} \bigg) \Bigg),\\%
C_1^{\text{SM}} =& 8 b \Bigg( \Delta a_{tu} \left(
\modulus{\left(F_V^+\right)_{\text{SM}}}^2 +
\modulus{\left(F_A^+\right)_{\text{SM}}}^2 \right) \nonumber\\%
& \qquad + \left( \Delta m^2 \right)_{12} \bigg( \Re\left(
\left(F_V^+\right)_{\text{SM}} \left(F_V^-\right)_{\text{SM}}^* \right)
\nonumber\\%
& \hspace{2.5cm} + \Re\left( \left(F_A^+\right)_{\text{SM}}
\left(F_A^-\right)_{\text{SM}}^* \right) \bigg) \Bigg),\\%
C_2^{\text{SM}} =& - 8 b^2 \left( \modulus{\left(F_V^+\right)_{\text{SM}}}^2 +
\modulus{\left(F_A^+\right)_{\text{SM}}}^2 \right).
\end{align}
\end{subequations}

\begin{figure}[hbtp]
\centering%
\includegraphics[scale=0.8]{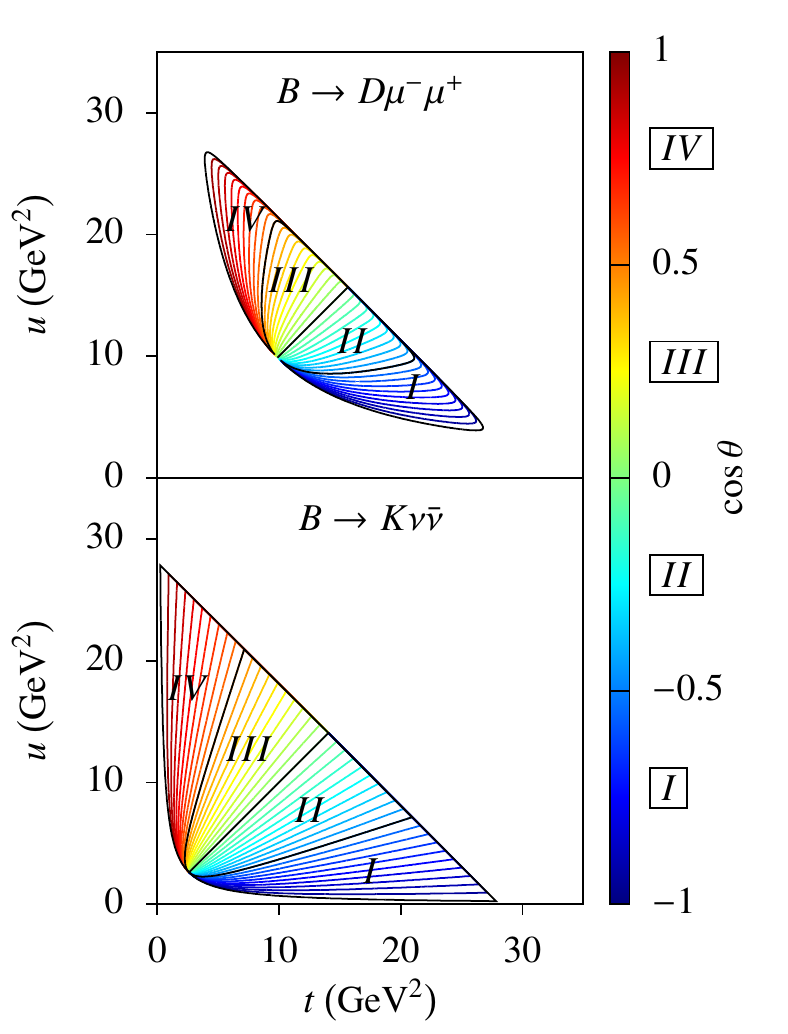}%
\caption{Two examples depicting the variation of $\cos\theta$ in the interior
	region of the $t$-vs-$u$ Dalitz plot. The interior of the Dalitz plot can be
	divided into four segments, $I$, $II$, $III$ and $IV$, as shown here.}%
\label{fig:Dalitz-plot-region}
\end{figure}

It is interesting to note that in the special case of $m_1 = m_2$, such as in
$P_i \to P_f \ell^+ \ell^-$, we always have $C_1^{\text{SM}}=0$. For specific
meson decays of the form $P_i \to P_f f_1 f_2$ allowed in the SM, one can write
down $\left(F_S\right)_{\text{SM}}$ , $\left(F_V^{\pm}\right)_{\text{SM}}$ and
$\left(F_A^{\pm}\right)_{\text{SM}}$, at least in principle. The SM prediction
for the angular distribution can thus be compared with corresponding
experimental measurement. In order to quantitatively compare the theoretical
prediction with experimental measurement, we define the following three angular
asymmetries which can precisely probe $C_0$, $C_1$ and $C_2$ individually,
\begin{subequations}\label{eq:ang-asymmetries}
\begin{align}
A_0 \equiv A_0(s) &= \frac{- \frac{1}{6} \left( \int_{-1}^{-1/2} - 7
	\int_{-1/2}^{+1/2} + \int_{+1/2}^{+1} \right) \dfrac{d^2 \Gamma}{ds \,
		d\cos\theta} d\cos\theta}{d\Gamma/ds} \nonumber\\%
&= 3C_0/\left(6C_0+2C_2\right),\\%
A_1 \equiv A_1(s) &= \frac{- \left( \int_{-1}^{0} - \int_{0}^{+1} \right)
	\dfrac{d^2 \Gamma}{ds \, d\cos\theta} d\cos\theta}{d\Gamma/ds} \nonumber\\%
&= 3C_1/\left(6C_0+2C_2\right),\\%
A_2 \equiv A_2(s) &= \frac{2 \left( \int_{-1}^{-1/2} - \int_{-1/2}^{+1/2} +
	\int_{+1/2}^{+1} \right) \dfrac{d^2 \Gamma}{ds \, d\cos\theta}
	d\cos\theta}{d\Gamma/ds} \nonumber\\%
&= 3C_2/\left(6C_0+2C_2\right).
\end{align}
\end{subequations}
The angular asymmetries of Eq.~\eqref{eq:ang-asymmetries} are functions of $s$
and it is easy to show that $A_2 = 3 \left(1/2 - A_0 \right)$. We can do the
integration over $s$ in Eq.~\eqref{eq:gen-angular-dist} and define the following
normalized angular distribution,
\begin{equation}\label{eq:Gen-ang-dist}
\frac{1}{\Gamma} \frac{d\Gamma}{d\cos\theta} = T_0 + T_1 \cos\theta + T_2 \cos^2\theta,
\end{equation}
where
\begin{equation}\label{eq:Def-T012}
T_j = 3 c_j/\left(6c_0 + 2 c_2\right),
\end{equation}
for $j=0,1,2$ and with
\begin{equation}\label{eq:cj}
c_j = \int_{(m_1+m_2)^2}^{(m_i-m_f)^2} \frac{b\sqrt{s} \, C_j}{128 \pi^3 m_i^2 \left(m_i^2 - m_f^2 + s\right)} ds.
\end{equation}
From Eq.~\eqref{eq:Def-T012} it is easy to show that $T_2 = 3 \left(1/2 -
T_0\right)$ which also ensures that integration over $\cos\theta$ on
Eq.~\eqref{eq:Gen-ang-dist} is equal to $1$. It is interesting to note that the
angular distribution of Eq.~\eqref{eq:Gen-ang-dist} can be written in terms of
the orthogonal Legendre polynomials of $\cos\theta$ as well,
\begin{equation}\label{eq:Gen-ang-dist-Legendre}
\frac{1}{\Gamma} \frac{d\Gamma}{d\cos\theta} = \sum_{i=0}^{2} \langle G^{(i)} \rangle P_i \left(\cos\theta\right).
\end{equation}
Here we have followed the notation of Ref.~\cite{Gratrex:2015hna} which also
analyzes decays of the type $P_i \to P_f f_1 f_2$, with only leptons for
$f_{1,2}$, in a model-independent manner but using a generalized helicity
amplitude method. The observables $\langle G^{(i)} \rangle$ of
Eq.~\eqref{eq:Gen-ang-dist-Legendre} are related to $T_0$, $T_1$ and $T_2$ of
Eq.~\eqref{eq:Gen-ang-dist} as follows,
\begin{subequations}
\begin{align}
\langle G^{(0)} \rangle &= T_0 + T_2/3 = 1/2,\\%
\langle G^{(1)} \rangle &= T_1,\\%
\langle G^{(2)} \rangle &= 2 T_2/3.
\end{align}
\end{subequations}
These angular observables $\langle G^{(i)} \rangle$'s can be obtained by using
the method of moments \cite{Gratrex:2015hna, Beaujean:2015xea}. Another
important way to describe the normalized angular distribution is by using a flat
term $F_H/2$ and the forward-backward asymmetry $A_{FB}$ \cite{AngDist:Hiller}
as follows,
\begin{equation}\label{eq:Gen-ang-dist-expt}
\frac{1}{\Gamma} \frac{d\Gamma}{d\cos\theta} = \frac{1}{2} F_H + A_{FB}
\cos\theta + \frac{3}{4} \left(1-F_H\right) \left(1 - \cos^2\theta\right).
\end{equation} 
This form of the angular distribution has also been used in the experimental
community \cite{AngDist:Expt} in the study of $B \to K \ell^+ \ell^-$. The parameters $F_H$ and $A_{FB}$ are related to $T_0$, $T_1$ and $T_2$ as follows,
\begin{subequations}
\begin{align}
F_H &= 2 \left( T_0 + T_2 \right) = 3 - 4 T_0,\\%
A_{FB} &= T_1.
\end{align}
\end{subequations}
Thus we have shown that Eqs.~\eqref{eq:Gen-ang-dist},
\eqref{eq:Gen-ang-dist-Legendre} and \eqref{eq:Gen-ang-dist-expt} are equivalent
to one another. In this paper, we choose to work using the normalized angular
distribution in terms of $T_0$, $T_1$ and $T_2$ as shown in
Eq.~\eqref{eq:Gen-ang-dist}. This is because the terms $T_0$, $T_1$ and $T_2$
can be easily determined experimentally by using the $t$-vs-$u$ Dalitz plot
which does not depend on any specific frame of reference. This Dalitz plot can
be easily divided into four segments $I$, $II$, $III$ and $IV$ as shown in
Fig.~\ref{fig:Dalitz-plot-region}. The segments are decided as follows,
\begin{center}
\begin{tabular}{lcl}
Segment $I$ & : & $-1 \leqslant \cos\theta \leqslant -0.5$,\\%
Segment $II$ & : & $-0.5 < \cos\theta \leqslant 0$,\\%
Segment $II$ & : & $0 < \cos\theta \leqslant 0.5$,\\%
Segment $IV$ & : & $0.5 < \cos\theta \leqslant 1$.%
\end{tabular}
\end{center}
The terms $T_0$, $T_1$ and $T_2$ can thus be expressed in terms of the following
asymmetries,
\begin{subequations}\label{eq:T012}
\begin{align}
T_0 &= - \frac{1}{6} \left( \frac{N_I - 7 \left( N_{II} + N_{III}\right) +
	N_{IV}}{N_I + N_{II} + N_{III} + N_{IV}} \right),\\%
T_1 &= \frac{\left( N_I + N_{II} \right) - \left( N_{III} + N_{IV} \right)}{N_I
	+ N_{II} + N_{III} + N_{IV}},\\%
T_2 &= 2 \left( \frac{N_I - \left( N_{II} + N_{III}\right) + N_{IV}}{N_I +
	N_{II} + N_{III} + N_{IV}} \right),
\end{align}
\end{subequations}
where $N_j$ denotes the number of events contained in the segment $j$. Since the
$t$-vs-$u$ Dalitz plot does not depend on the frame of reference, we need not
constraint ourselves to the Gottfried-Jackson frame of Fig.~\ref{fig:GJ-frame}
and can work in the laboratory frame as well. Furthermore, we can use the
expressions in Eq.~\eqref{eq:T012} to search for NP.

\section{Illustrating the effects of new physics on the angular distribution}\label{sec:example}

\subsection{Classification of the \texorpdfstring{$P_i \to P_f f_1 f_2$}{Pi -> Pf + f1 + f2} decays}%
It should be emphasized that for our methodology to work, we need to know the
angle $\theta$ in the Gottfried-Jackson frame, or equivalently the $t$-vs-$u$
Dalitz plot, which demand that 4-momenta of the final particles be fully known.
Usually, the 4-momenta of the initial and final pseudo-scalar mesons are
directly measured experimentally. However, depending on the detection
possibilities of $f_1$ and $f_2$ we can identify three distinct scenarios for
our process $P_i \to P_f f_1 f_2$. We introduce the notations $f_i^{\CheckMark}$
and $f_i^{\CrossMark}$ to denote whether the fermion $f_i$ gets detected
(\CheckMark) or not (\CrossMark) by the detector. Using this notation the three
scenarios are described as follows.
\begin{enumerate}
\item[(S1)] $P_i \to P_f + f_1^{\CheckMark} + f_2^{\CheckMark} \equiv P_f +
\textrm{`visible'}$. Here both $f_1$ and $f_2$ are detected, e.g.\ when $f_1 f_2
= \ell^-\ell^+$ or $\ell \bar{\ell'}$.%
\item[(S2)] $P_i \to
\begin{Bmatrix}
P_f + f_1^{\CheckMark} + f_2^{\CrossMark}\\%
P_f + f_1^{\CrossMark} + f_2^{\CheckMark}
\end{Bmatrix} \equiv P_f + \textrm{`visible'} + \text{`invisible'}$. Here either
$f_1$ or $f_2$ gets detected, e.g.\ when $f_1 f_2 = \ell \nu_{\ell}$, $\ell
\nu_S$, $\ell f^{DM}$, $\ell f^{LLP}$.%
\item[(S3)] $P_i \to P_f + f_1^{\CrossMark} + f_2^{\CrossMark} \equiv P_f +
\textrm{`invisible'}$. Here neither $f_1$ nor $f_2$ gets detected, e.g.\ when
$f_1 f_2 = \nu_{\ell}\nubar_{\ell}$, $\nu_{\ell}\nubar_S$, $\nu_S\nubar_{\ell}$,
$\nu_S\nubar_S$, $f^{DM} \bar{f}^{DM}$, $f_1^{DM} f_2^{DM}$, $f_1^{LLP}
f_2^{LLP}$ etc.
\end{enumerate}
It should be noted that the above classification is based on our existing
experimental explorations. What is undetected today might get detected in future
with advanced detectors. In such a case we can imagine that, in future, the
modes grouped in S2 might migrate to S1 and those in S3 might be grouped under
S2. Below we explore each of the above scenarios in more details.

\subsection{Exploration of new physics effects in each scenario}

The first scenario (S1) is an experimenter's delight as in this case all final
4-momenta can be easily measured and the $t$-vs-$u$ Dalitz plot can be obtained.
Here, our methodology can be used to look for the possible signature of new
physics in rare decays such as $B \to D \ell^- \ell^+$ (which can be found in
\cite{Kim:2016zbg}) or study the nature of new physics contributing to
lepton-flavor violating processes such as $B \to P \ell^{\pm} \ell^{\prime\mp}$
where $P=\pi,K,D$, $\ell\neq\ell'$ and $\ell,\ell'=e,\mu,\tau$. Let us consider
a few NP possibilities mediating this lepton-flavor violating decay. There is no
contribution within the SM to such decays. Therefore, all contribution to these
decays comes from NP alone. It is very easy to note that for the decay $B \to P
\ell^{-} \ell^{\prime+}$, from Eqs.~\eqref{eq:gen-C012} and
\eqref{eq:Gen-ang-dist} we get,
\begin{equation}\label{eq:NPinB2PLl}
\frac{1}{\Gamma} \frac{d\Gamma}{d\cos\theta} = 
\begin{cases}
\dfrac{1}{2}, & \left(\parbox{0.23\linewidth}{\textrm{\centering only scalar or pseudo-scalar interaction}}\right)\\[5mm]%
T_0 + T_2 \cos^2\theta, & \left(\parbox{0.23\linewidth}{\textrm{\centering only tensorial interaction}}\right)\\[3mm]%
T_0 + T_1 \cos\theta + T_2 \cos^2\theta, & \left(\parbox{0.23\linewidth}{\textrm{\centering only vector or axial-vector interaction}}\right)
\end{cases}
\end{equation}
where $T_2 = 3\left(1/2 - T_0\right)$ with the quantities $T_0$, $T_1$ and $T_2$
being easily obtainable from the Dalitz plot distribution by using
Eq.~\eqref{eq:T012}. It is clear from Eq.~\eqref{eq:NPinB2PLl} that scalar or
pseudo-scalar interaction would give rise to a uniform (or constant) angular
distribution, while tensorial interaction gives a non-uniform distribution which
is symmetric under $\cos\theta \leftrightarrow -\cos\theta$ and for this $T_0
\leqslant 1/2$. On the other hand vector or axial-vector interaction can only be
described by the most general form of the angular distribution, with its
signature being $T_1 \neq 0$. Nevertheless, if vector or axial-vector
interaction contributes to the flavor violating processes $B \to P \ell^{-}
\ell^{\prime+}$, it is important to note that $T_1 \propto \left(m_{\ell}^2 -
m_{\ell'}^2\right)$, where $m_{\ell}$, $m_{\ell'}$ denote the masses of the
charged leptons $\ell^-$ and $\ell^{\prime+}$ respectively. Therefore, we should
observe an increase in the value of $T_1$ when going from $B \to P \mu^- e^+$ to
$B \to P \tau^- \mu^+$ to $B \to P \tau^- e^+$. This would nail down the vector
or axial vector nature of the NP, if it is the only NP contributing to these
decays. Thus far we have analyzed the first scenario (S1) in which the relevant
decays can be easily probed with existing detectors.

The second scenario (S2) can also be studied experimentally with existing
detectors. In this case, the missing 4-momentum can be fully deduced using
conservation of 4-momentum. Thus the $t$-vs-$u$ Dalitz plot can readily be
obtained. Using our methodology the signatures of NP can then be extracted. One
promising candidate for search for NP in this kind of scenario is in the decay
$B \to P \ell N$ where $P=\pi$, $K$ or $D$ and $N$ can be an active neutrino
($\nu_{\ell}$) or sterile neutrino ($\nu_S$) or a neutral dark fermion
($f^{DM}$) or a long lived neutral fermion ($f^{LLP}$) which decays outside the
detector. These S2 decay modes offer an exciting opportunity for study of NP
effects.

The third scenario (S3), which has the maximum number of NP possibilities, is
also the most challenging one for the current generation of experimental
facilities, due to lack of information about the individual 4-momentum of $f_1$
and $f_2$. This implies that we can not do any angular analysis for these kind
of decays unless by some technological advancement such as by using displaced
vertex detectors\footnote{There are many existing proposals for such displaced
	vertex studies from other theoretical and experimental considerations (see
	Refs.~\cite{DV:Theory,DV:Experiments} and references contained therein for
	further information). } we can manage to make measurement of the 4-momentum or
the angular information of at least one of the final fermions. Getting 4-momenta
of both the fermions would be ideal, but knowing 4-momentum of either one of
them would suffice for our purpose. We are optimistic that the advancement in
detector technology would push the current S3 decay modes to get labelled as S2
modes in the foreseeable future. It is important to note that once the current
S3 modes enter the S2 category, we can cover the whole spectrum of NP
possibilities in the $P_i \to P_f f_1 f_2$ decays. Below we make a comprehensive
exploration of NP possibilities in the generalized S2 decay modes, which
includes the current S2 and S3 modes together.

\subsection{Probing effects of new physics in the (S2)\\and generalized (S2)	scenarios}

In the generalized S2 (GS2) scenario we have decays of the type $P_i \to
\begin{Bmatrix}
P_f + f_1^{\CheckMark} + f_2^{\CrossMark}\\%
P_f + f_1^{\CrossMark} + f_2^{\CheckMark}
\end{Bmatrix} \equiv P_f + \textrm{`visible'} + \text{`invisible'}$, where the detected (\CheckMark) or undetected (\CrossMark) nature is not constrained by our existing detector technology. In some cases, even with advanced detectors, either of the fermions $f_1$, $f_2$ might not get detected simply because its direction of flight lies outside the finite detector coverage, especially when the detector is located farther from the place of origin of the particle. Such possibilities are also included here. As noted before, measuring the 4-momentum of either of the final fermions would suffice to carry out the angular analysis following our approach. 

In this context let us analyze the following decays.
\begin{enumerate}
\item[(i)] S2 decay: $B \to P\ell^- f^{\CrossMark}$ where $P$ can be $\pi$ or
$D$ and $f^{\CrossMark}$ is a neutral fermion. In the SM this process is
mediated by $W^-$ boson and we have $f^{\CrossMark} = \nubar_{\ell}$. Presence
of NP can imply $f^{\CrossMark}$ being a sterile neutrino $\nu_S$ or a fermionic
dark matter particle $f^{DM}$ or a long lived fermion $f^{LLP}$, with additional
non-SM interactions.%
\item[(ii)] GS2 decay: $B \to K f_1^{\CheckMark} f_2^{\CrossMark}$ where
$f_1^{\CheckMark}$ and $f_2^{\CrossMark}$ are both neutral fermions. In the SM
this process is mediated by $Z^0$ boson and we have $f_1 f_2 = \nu_{\ell}
\nubar_{\ell}$. However, in case of NP contribution we can get pairs of sterile
neutrinos or fermionic dark matter or fermionic long lived particles etc.\ along
with nonstandard interactions as well. Here we are assuming that either of the
final neutral fermions leaves a displaced vertex signature in an advanced
detector so that its 4-momentum or angular information could be obtained.%
\end{enumerate}

\subsubsection{New physics effects in the S2 decay \texorpdfstring{$B \to P\ell^- f^{\CrossMark}$}{B -> P + l- + fX}}
Analyzing the $B \to P\ell^- f^{\CrossMark}$ decay in the SM we find that only
vector and axial vector currents contribute and $F_A^{\pm} = -F_V^{\pm}$ while
other form factors are zero. Also considering the anti-neutrino to be massless,
i.e.\ $m_2 =0$ we find that 
\begin{align*}
a_t &= m_{\ell}^2 + m_P^2 + \left(s + m_{\ell}^2\right) \left(m_B^2 - m_P^2 -
s\right)/(2s),\\%
a_u &= m_P^2 + \left(s - m_{\ell}^2\right) \left(m_B^2 - m_P^2 -
s\right)/(2s),\\%
b &= \left(s-m_{\ell}^2\right) \sqrt{\lambda\left(s, m_B^2, m_P^2\right)}/(2s),
\end{align*}
where $m_{\ell}$, $m_P$ and $m_B$ denote the masses of the charged lepton
$\ell^-$, mesons $P$ and $B$ respectively. Substituting these information in
Eqs.~\eqref{eq:SM-C012} and in Eq.~\eqref{eq:gen-angular-dist} we get,
\begin{equation}\label{eq:B2Plnu-dist-gen}
\frac{d^2\Gamma^{\textrm{SM}}}{ds \, d\cos\theta} = \frac{b\sqrt{s} \left(
	C_0^{\textrm{SM}} + C_1^{\textrm{SM}} \cos\theta + C_2^{\textrm{SM}}
	\cos^2\theta \right)}{128 \, \pi^3 \, m_B^2 \left(m_B^2 - m_P^2 + s \right)},
\end{equation}
where
\begin{subequations}\label{eq:C012-in-B2Plnu}
\begin{align}
C_0^{\text{SM}} =& 4 \Bigg( \modulus{\left(F_V^+\right)_{\text{SM}}}^2 \bigg(
\lambda\left(s, m_B^2, m_P^2\right) - m_{\ell}^2 \left(s - 2 \left(m_B^2 -
m_P^2\right) \right) \nonumber\\%
& \hspace{2cm} - m_{\ell}^4 \left(m_B^2 - m_P^2\right)^2/s^2 \bigg)
\nonumber\\%
& \quad + \modulus{\left(F_V^-\right)_{\text{SM}}}^2 m_{\ell}^2 \left( s -
m_{\ell}^2 \right) \nonumber\\%
& \quad + 2 \Re\left( \left(F_V^+\right)_{\text{SM}}
\left(F_V^-\right)_{\text{SM}}^* \right) m_{\ell}^2 \left(m_B^2 - m_P^2\right)
\left(1- \frac{m_{\ell}^2}{s}\right) \Bigg),\\%
C_1^{\text{SM}} =& 16 m_{\ell}^2 b \Bigg( \left(\frac{m_B^2 - m_P^2}{s}\right)
\modulus{\left(F_V^+\right)_{\text{SM}}}^2 + \Re\left(
\left(F_V^+\right)_{\text{SM}} \left(F_V^-\right)_{\text{SM}}^* \right)
\Bigg),\\%
C_2^{\text{SM}} =& - 16 b^2 \modulus{\left(F_V^+\right)_{\text{SM}}}^2 .
\end{align}
\end{subequations}
It is important to notice that in Eq.~\eqref{eq:C012-in-B2Plnu} we have many
terms in the expression for $C_0^{\textrm{SM}}$ that are proportional to some
power of the lepton mass, while the entire $C_1^{\textrm{SM}}$ is directly
proportional to $m_{\ell}^2$. If we compare the $m_{\ell}$ dependent and
$m_{\ell}$ independent contributions in $C_0^{\textrm{SM}}$ we find that the
dependent terms are suppressed by about a factor of
$\mathcal{O}\left(2m_{\ell}^2/m_B^2\right)$ which is roughly $8\times 10^{-4}$
for muon and $2\times 10^{-8}$ for electron. Thus we can neglect these
$m_{\ell}$ dependent terms in comparison with mass independent terms.
Equivalently, we can consider the charged leptons such as electron and muon as
massless fermions, when compared with the $B$ meson mass scale. In the limit
$m_{\ell} \to 0$ the expression for angular distribution as given in
Eq.~\eqref{eq:B2Plnu-dist-gen} becomes much simpler,
\begin{equation}
\frac{d^2\Gamma^{\text{SM}}}{ds \, d\cos\theta} = \frac{b^3\sqrt{s}}{8 \, \pi^3
	\, m_B^2 \left( m_B^2 - m_P^2 + s \right)}
\modulus{\left(F_V^+\right)_{\text{SM}}}^2 \sin^2\theta.
\end{equation}
Independent of the expression for $\left(F_V^+\right)_{\text{SM}}$, it is easy
to show that the normalized angular distribution is given by,
\begin{equation}\label{eq:SM-Dist-B2Plnu-massless}
\frac{1}{\Gamma^{\text{SM}}} \frac{d\Gamma^{\text{SM}}
}{d\cos\theta} = \frac{3}{4} \sin^2\theta,
\end{equation}
which implies that $T_0 = 3/4 = -T_2$, $T_1 = 0$. Since the distribution of
events in the Dalitz plot is symmetric under $\cos\theta \leftrightarrow -
\cos\theta$, we have $N_I = N_{IV}$ and $N_{II} = N_{III}$ which automatically
satisfies the condition $T_1 = 0$. If we solve $T_0 = 3/4 = -T_2$, we find that
the number of events in the different segments of the Dalitz plot (equivalently
the number of events in the four distinct bins of $\cos\theta$) are related to
one another by
\begin{equation}\label{eq:SM-bins-B2Plnu}
\frac{N_I}{N_{II}} = \frac{5}{11} = \frac{N_{IV}}{N_{III}}.
\end{equation}
Any significant deviation from this would imply presence of NP effects. To
illustrate the effects of NP on the angular distribution in these types of
decays, we consider two simple and specific NP possibilities. Here we assume the
charged lepton to be massless ($m_{\ell}=0$) and the undetected fermion
($f^{\CrossMark}$) to have mass $m\neq 0$.

\begin{figure}[hbtp]
\centering%
\includegraphics[scale=0.8]{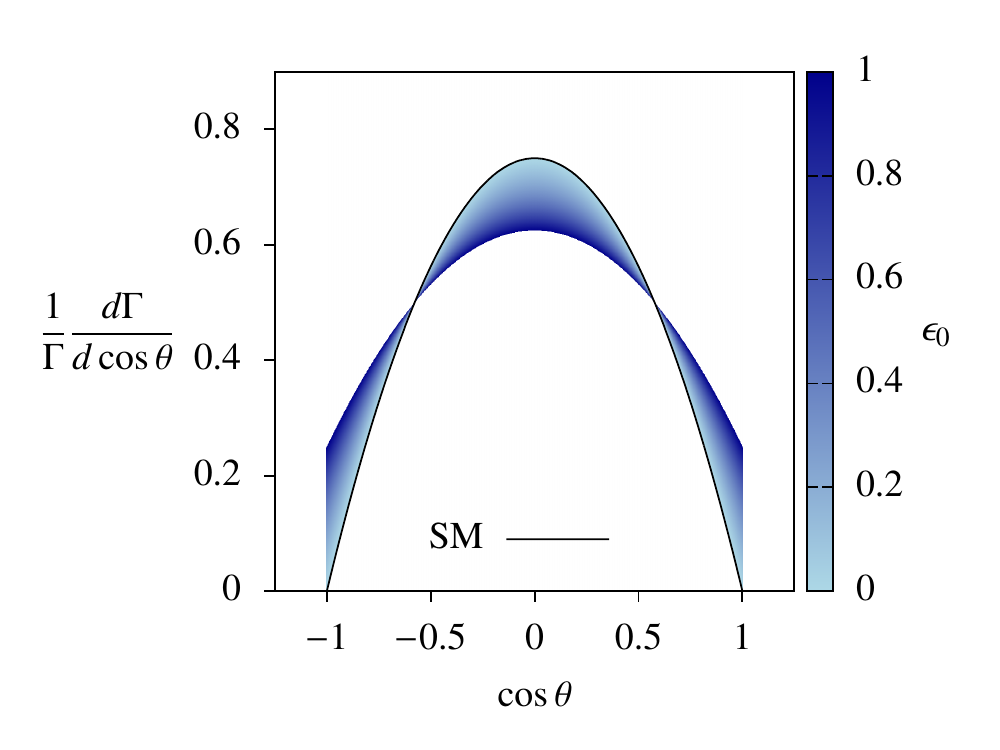}%
\caption{Normalized uni-angular distribution showing the effect of a scalar new
	physics contribution to $B \to P \ell^- f^{\CrossMark}$ where we have neglected
	the mass of the charged lepton $\ell =e,\mu$. This also shows the normalized
	uni-angular distribution showing the effect of a scalar new physics contribution
	to $B \to K f_1^{\CheckMark} f_2^{\CrossMark}$ considering the $m_1 = m_2$ case
	only.}%
\label{fig:Scalar-NP-B2Plnu}
\end{figure}

\paragraph{\textbf{Scalar type new physics:}} Considering the simplest scalar type NP
scenario, with $F_S \neq 0$, $F_P = F_V^{\pm} = F_A^{\pm} = F_{T_1} = F_{T_2} =
0$, we get
\begin{align*}
C_0^{\text{NP}} =& 2 \left(s - m^2\right) \modulus{F_S}^2,\\%
C_1^{\text{NP}} =& 0 = C_2^{\text{NP}}.
\end{align*}
In other words, there is no angular dependence at all here, i.e.\
\begin{equation*}
\frac{d^2\Gamma^{\text{NP}}}{ds \, d\cos\theta} = \frac{b\sqrt{s}}{64 \, \pi^3 \,
	m_B^2 \left(m_B^2 - m_P^2 + s \right)} \left(s - m^2\right) \modulus{F_S}^2,
\end{equation*}
where $b = \left(s-m^2\right) \sqrt{\lambda\left(s,m_B^2,m_P^2\right)}/(2s)$ and
$m^2 \leqslant s \leqslant \left(m_B-m_P\right)^2$. If we do the integration
over $s$, then the normalized angular distribution is given by,
\begin{equation*}
\frac{1}{\Gamma^{\text{NP}}} \frac{d\Gamma^{\text{NP}}}{d\cos\theta} =
\frac{1}{2}.
\end{equation*}
In fact, if such a new physics were present, our observation of $B \to P +
\ell^- + f^{\CrossMark}$ would have the following angular distribution,
\begin{equation*}
\frac{d\Gamma}{d\cos\theta} = \Gamma^{\text{SM}} \left(\frac{3}{4} \sin^2\theta
+ \frac{1}{2} \epsilon_0 \right),
\end{equation*}
where we have parametrized the new physics contribution in terms of $\epsilon_0$,
\begin{equation*}
\epsilon_0 = \Gamma^{\text{NP}}/\Gamma^{\text{SM}}.
\end{equation*}
Doing integration over $\cos\theta$ we get,
\begin{equation*}
\Gamma = \Gamma^{\text{SM}} \left(1+\epsilon\right) = \Gamma^{\text{SM}} + \Gamma^{\text{NP}}.
\end{equation*}
This implies 
\begin{equation}\label{eq:Scalar-NP-Dist-B2Plnu}
\frac{1}{\Gamma} \frac{d\Gamma}{d\cos\theta} = \frac{3\sin^2\theta + 2
	\epsilon_0}{4 \left(1+\epsilon_0\right)}.
\end{equation}
This angular distribution is shown in Fig.~\ref{fig:Scalar-NP-B2Plnu} where we
have varied $\epsilon_0$ in the range $[0,1]$, i.e.\ we have allowed the
possibility that the NP contribution might be as large as that of the SM. It is
interesting to find that in Fig.~\ref{fig:Scalar-NP-B2Plnu} at two specific
values of $\cos\theta$ there is no difference between the standard model
prediction alone and the combination of standard model and new physics
contributions. These two points can be easily obtained by equating
Eqs.~\eqref{eq:SM-Dist-B2Plnu-massless} and \eqref{eq:Scalar-NP-Dist-B2Plnu},
and then solving for $\cos\theta$ gives us
\begin{equation}
\cos\theta = \pm 1/\sqrt{3} \approx \pm 0.57735.
\end{equation}
This corresponds to the angle $\theta \approx 54.74^{\circ}$. At these two
points in $\cos\theta$, the normalized uni-angular distribution always has the
value $0.5$, even if there is some scalar new physics contributing to our
process under consideration.

From Eq.~\eqref{eq:Scalar-NP-Dist-B2Plnu} it is clear that despite the scalar NP
effect, the distribution is still symmetric under $\cos\theta \leftrightarrow
-\cos\theta$, and solving for the number of events in the four segments of the
Dalitz plot (equivalently the four $\cos\theta$ bins) we get,
\begin{equation}\label{eq:Scalar-NP}
\frac{N_I}{N_{II}} = \frac{5+8\epsilon_0}{11 + 8\epsilon_0} =
\frac{N_{IV}}{N_{III}}.
\end{equation}
It is easy to see that when $\epsilon=0$ we get back the SM prediction of
Eq.~\eqref{eq:SM-bins-B2Plnu} as expected.

\paragraph{\textbf{Tensor type new physics:}} 

\begin{figure*}[hbtp]
\centering%
\includegraphics[scale=0.8]{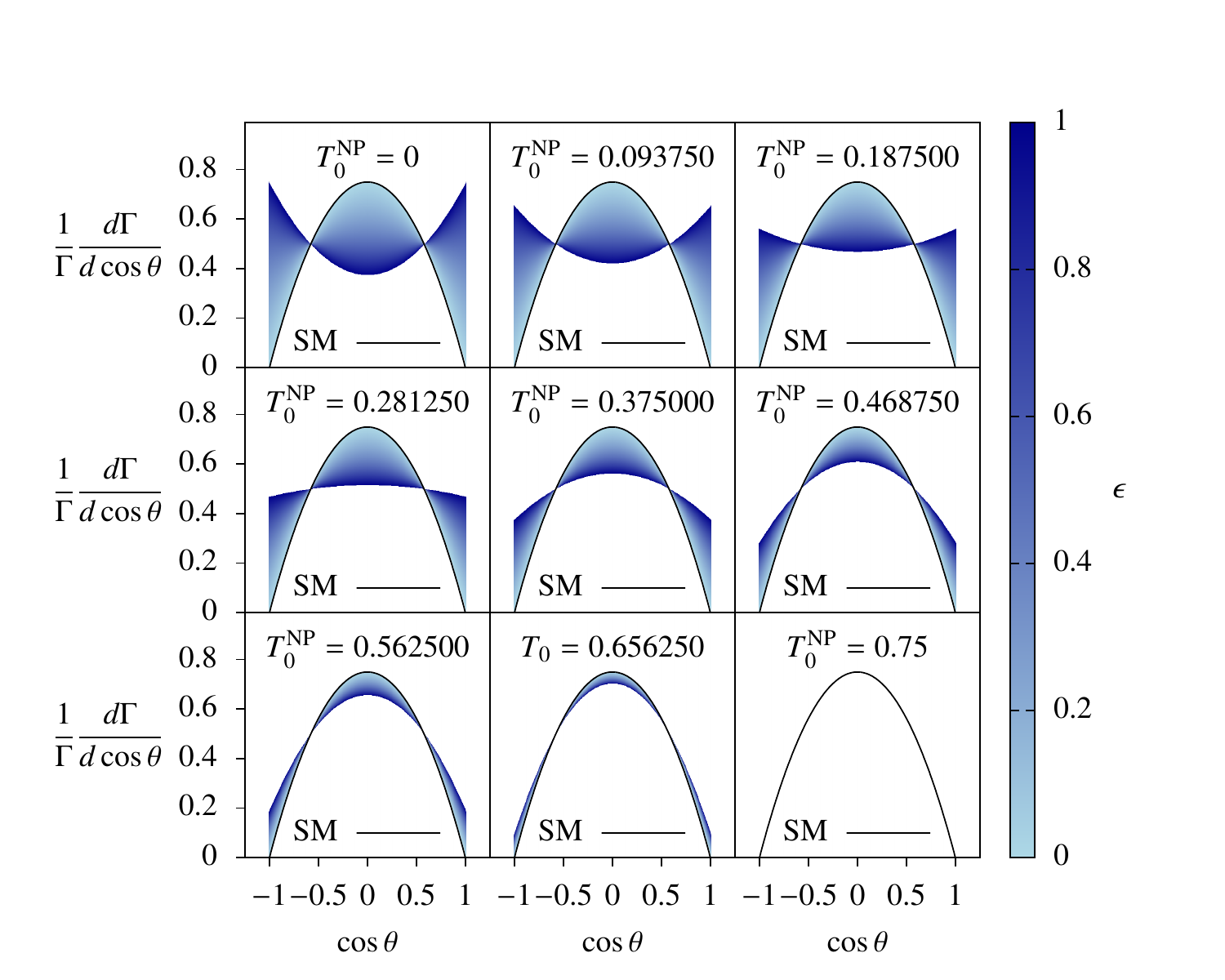}%
\caption{Normalized uni-angular distribution showing the effect of a tensor new
	physics contribution to $B \to P \ell^- f^{\CrossMark}$ where we have neglected
	the mass of the charged lepton $\ell=e,\mu$. These set of plots can also
	describe the effect of a vector new physics contribution to $B \to K
	f_1^{\CheckMark} f_2^{\CrossMark}$ when the final fermions are equally
	massive.}%
\label{fig:Tensor-NP}
\end{figure*}

Let us consider a tensor type of new physics
possibility in which $F_{T_1} \neq 0$ and all other form factors are zero. In
such a case we get,
\begin{align*}
C_0^{\textrm{NP}} &=2 m^2 \left(s-m^2\right)
\frac{\lambda\left(s,m_B^2,m_P^2\right)}{s} \modulus{F_{T_1}}^2,\\%
C_1^{\textrm{NP}} &=0,\\%
C_2^{\textrm{NP}} &= 2 \left(s-m^2\right)^2 \frac{\lambda\left(s, m_B^2,
	m_P^2\right)}{s} \modulus{F_{T_1}}^2.
\end{align*}
It is easy to notice that in the limit $m \to 0$ we have $C_0 \to 0$ but $C_2
\not\to 0$. If we do the integration over $s$, then the normalized angular
distribution is given by,
\begin{equation}
\frac{1}{\Gamma^{\textrm{NP}}} \frac{d\Gamma^{\textrm{NP}}}{d\cos\theta} =
T_0^{\textrm{NP}} + T_2^{\textrm{NP}} \cos^2\theta,
\end{equation}
where $T_2^{\textrm{NP}} = 3\left(1/2-T_0^{\textrm{NP}}\right)$ and
$T_0^{\textrm{NP}} = 3c_0/\left(6c_0 + 2c_2\right)$ with
\begin{equation*}
c_j = \int_{m^2}^{\left(m_B-m_P\right)^2} \frac{b\sqrt{s} \
	C_j^{\textrm{NP}}}{128 \pi^3 m_B^2 \left(m_B^2 - m_P^2 + s\right)} ds.
\end{equation*}
Thus in the limit $m\to 0$ we have $T_0 = 0$. If such a new physics were
present, our observation of $B \to P \ell^- f^{\CrossMark}$ would have the
following angular distribution,
\begin{equation}
\frac{d\Gamma}{d\cos\theta} = \Gamma^{\text{SM}} \left(\frac{3}{4} \sin^2\theta + \left(T_0^{\textrm{NP}} + 3 \left(\frac{1}{2} - T_0^{\textrm{NP}}\right) \cos^2\theta\right) \epsilon \right),
\end{equation}
where $\epsilon=\Gamma^{\textrm{NP}}/\Gamma^{\textrm{SM}}$ is the NP parameter
which can vary in the range $\left[0,1\right]$ denoting the possibility that the
NP contribution can be as large as that of the SM, and $T_0^{\textrm{NP}}$ acts
as a free parameter here which can vary in the range $\left[0,3/4\right]$ in
which $d\Gamma^{\textrm{NP}}/d\cos\theta \geqslant 0$ for all values of
$\cos\theta$. Doing integration over $\cos\theta$ we get $\Gamma =
\Gamma^{\textrm{SM}} \left(1 + \epsilon\right) = \Gamma^{\textrm{SM}} +
\Gamma^{\textrm{NP}}$. This implies
\begin{equation}\label{eq:Tensor-NP-Dist-B2Plnu}
\frac{1}{\Gamma} \frac{d\Gamma}{d\cos\theta} = \frac{3 + 4 T_0^{\textrm{NP}}
	\epsilon - 3 \left(4T_0^{\textrm{NP}} \epsilon -2\epsilon +
	1\right)\cos^2\theta}{4\left(1+\epsilon\right)}.
\end{equation}

This angular distribution is shown in Fig.~\ref{fig:Tensor-NP} in which we have
considered nine values of $T_0^{\textrm{NP}}$ and varied $\epsilon$ in the range
$[0,1]$. It is clearly evident in Fig.~\ref{fig:Tensor-NP} that
$T_0^{\textrm{NP}} = 3/4$ case is always indistinguishable from the SM case, as
it should be. Just like the scalar-type new physics case, we observe that there
are two values of $\cos\theta$ at which there is no difference between the SM
prediction alone and the combination of SM and NP contributions. These two
points can be easily computed by equating
Eqs.~\eqref{eq:SM-Dist-B2Plnu-massless} and \eqref{eq:Tensor-NP-Dist-B2Plnu},
and then solving for $\cos\theta$ we once again find that,
\begin{equation}
\cos\theta = \pm 1/\sqrt{3} \approx \pm 0.57735,
\end{equation}
which corresponds to the angle $\theta \approx 54.74^{\circ}$. At these two
points in $\cos\theta$, the normalized uni-angular distribution always has the
value $0.5$, even if there is some tensor new physics contributing to our
process under consideration. It should be noted that these are also the same
points where the scalar new physics contribution shows similar effect.

It is also easy to notice that the angular distribution as given in
Eq.~\eqref{eq:Tensor-NP-Dist-B2Plnu} is symmetric under $\cos\theta
\leftrightarrow -\cos\theta$, and solving for the number of events in the four
segments of the Dalitz plot (equivalently the four $\cos\theta$ bins) we get,
\begin{equation}
\frac{N_{I}}{N_{II}} = \frac{5 + 2 \epsilon \left(7 - 6
	T_0^{\textrm{NP}}\right)}{11 + 2 \epsilon \left(1 + 6 T_0^{\textrm{NP}}\right)}
= \frac{N_{IV}}{N_{III}}.
\end{equation}
It is easy to see that when $\epsilon=0$ or $T_0^{\textrm{NP}}=3/4$ we get back
the SM prediction of Eq.~\eqref{eq:SM-bins-B2Plnu} as expected.

Finally we analyze new physics possibilities in the decays belonging to the GS2
category. Due to the very nature of the GS2 decay modes, the following
discussion of NP effects presumes usage of advanced detector technology to get
angular information.

\subsubsection{New physics effects in the GS2 decay \texorpdfstring{$B \to K f_1^{\CheckMark} f_2^{\CrossMark}$}{B -> K + f1V + f2X}}

As mentioned before, the GS2 decay modes are originally part of S3, i.e.\ it is
extremely difficult to get angular distribution for these cases unless we
innovate on detector technology. Here we consider such a decay mode $B \to K
f_1^{\CheckMark} f_2^{\CrossMark}$ in which both $f_1$, $f_2$ are neutral
fermions who have evaded, till now, all our attempts to detect them near their
place of origin. But probably with displaced vertex detectors or some other
advanced detector we could bring at least one of these fermions (say $f_1$)
under the purview of experimental study and measure its 4-momentum or angular
information. The missing fermion (which is $f_2$ in our example here) might have
flied in a direction along which there is no detector coverage. To increase the
sample size we should include $B \to K f_1^{\CrossMark} f_2^{\CheckMark}$ events
also, provided we know how to ascertain the particle or anti-particle nature of
$f_1$ and $f_2$. To illustrate this point, let us consider the possibility $f_1
f_2 = \nu_S \nubar_S$. In a displaced vertex detector if we see $\pi^+ \mu^-$
events, they can be attributed to the decay of $\nu_S$ and similarly $\pi^-
\mu^+$ events would appear from the decay of $\nubar_S$. In this case, we can
infer the angle $\theta$ by knowing the 4-momentum of either $f_1 = \nu_S$ or
$f_2 = \nubar_S$ (see Fig.~\ref{fig:GJ-frame}). If we find that both $f_1$ and
$f_2$ leave behind their signature tracks in the detector (i.e.\
$f_1^{\CheckMark} f_2^{\CheckMark}$) it would be the most ideal situation. But
as we have already stressed before, measuring 4-momenta of either of the
fermions would suffice for our angular studies.

In the SM the only contribution to $B \to K f_1^{\CheckMark} f_2^{\CrossMark}$
and $B \to K f_1^{\CrossMark} f_2^{\CheckMark}$ would come from $B \to K
\nu_{\ell} \nubar_{\ell}$ where as in the case of NP we have a number of
possibilities that includes sterile neutrinos, dark matter particles, or some
long lived particles in the final state, $f_1 f_2 = \nu_{\ell} \nubar_S$, $\nu_S
\nubar_{\ell}$, $\nu_S \nubar_S$, $f^{\textrm{DM}} \bar{f}^{\textrm{DM}}$,
$f_1^{\textrm{DM}} f_2^{\textrm{DM}}$, $f^{\textrm{LLP}}
\bar{f}^{\textrm{LLP}}$, $f_1^{\textrm{LLP}} f_2^{\textrm{LLP}}$
etc.\footnote{In addition to the new physics possibilities considered here,
	there can be additional contributions to the $B \to K + \text{`invisible'}$
	decay, e.g.\ from SM singlet scalars contributing to the `invisible' part as
	discussed in Ref.~\cite{Kim:2009qc}. As is evident, our analysis is instead
	focused on a pair of fermions contributing to the `invisible' part.} One can
also consider non-standard neutrino interactions also contributing in these
cases. To demonstrate our methodology, we shall analyze only a subset of these
various NP possibilities in which $f_1$ and $f_2$ have the same mass, i.e.\ $m_1
= m_2 = m$ (say), as this greatly simplifies the calculation. As we shall
illustrate below we can not only detect the presence of NP but ascertain whether
it is of scalar type or vector type, for example, by analyzing the angular
distribution.

Before, we go for new physics contributions, let us analyze the SM contribution
$B \to K \nu_{\ell} \nubar_{\ell}$. Here only vector and axial-vector currents
contributions, and $F_A^{\pm} = - F_V^{\pm}$. Also the neutrino and
anti-neutrino are massless, i.e.\ $m_1 = 0 = m_2$, which implies $a_t = a_u =
\tfrac{1}{2} \left(m_B^2 + m_K^2 -s\right)$ and $b = \tfrac{1}{2}
\sqrt{\lambda\left(s,m_B^2,m_K^2\right)}$, where $m_B$ and $m_K$ denote the
masses of $B$ and $K$ mesons respectively. Substituting these information in
Eqs.~\eqref{eq:SM-C012} and in Eq.~\eqref{eq:gen-angular-dist} we get,
\begin{equation}
\frac{d^2\Gamma^{\text{SM}}}{ds \, d\cos\theta} = \frac{b^3\sqrt{s}}{8 \, \pi^3 \, m_B^2
	\left( m_B^2 - m_K^2 + s \right)} \modulus{\left(F_V^+\right)_{\text{SM}}}^2
\sin^2\theta.
\end{equation}
Irrespective of the expression for $\left(F_V^+\right)_{\text{SM}}$, it can be
easily shown that the normalized angular distribution is given by,
\begin{equation}\label{eq:SM-Dist-B2Knn}
\frac{1}{\Gamma^{\text{SM}}} \frac{d\Gamma^{\text{SM}}
}{d\cos\theta} = \frac{3}{4} \sin^2\theta,
\end{equation}
which implies that $T_0 = 3/4 = -T_2$, $T_1 = 0$. Following the same logic as
the one given after Eq.~\eqref{eq:SM-Dist-B2Plnu-massless}, we find that the
number of events in the different segments of the Dalitz plot (equivalently the
number of events in the four distinct bins of $\cos\theta$) are related to one
another by,
\begin{equation}\label{eq:SM-bins}
\frac{N_I}{N_{II}} = \frac{5}{11} = \frac{N_{IV}}{N_{III}}.
\end{equation}
This sets the stage for us to explore (i) a scalar type and (ii) a vector type
of NP possibility, with final fermions for which $m_1 = m_2 = m \neq 0$. 

\paragraph{\textbf{Scalar type new physics:}}

Once again we consider the simplest scalar type NP scenario, with $F_S \neq 0$,
and other form factors being zero. This leads us to,
\begin{align*}
	C_0^{\text{NP}} =& 2 \left(s - 4m^2\right) \modulus{F_S}^2,\\%
	C_1^{\text{NP}} =& 0 = C_2^{\text{NP}}.
\end{align*}
In other words, there is no angular dependence at all here, i.e.\
\begin{equation}
\frac{d^2\Gamma^{\text{NP}}}{ds \, d\cos\theta} = \frac{b\sqrt{s}}{64 \, \pi^3 \,
	m_B^2 \left(m_B^2 - m_K^2 + s \right)} \left(s - 4m^2\right) \modulus{F_S}^2,
\end{equation}
where $b = \left(\sqrt{\left(s-4m^2\right)} \,
\sqrt{\lambda\left(s,m_B^2,m_K^2\right)}\right)/(2\sqrt{s})$ and $4m^2 \leqslant
s \leqslant \left(m_B-m_K\right)^2$. If we do the integration over $s$, then for
NP only the normalized angular distribution is given by,
\begin{equation*}
\frac{1}{\Gamma^{\text{NP}}} \frac{d\Gamma^{\text{NP}}}{d\cos\theta} =
\frac{1}{2}.
\end{equation*}
Assuming such a NP contributing in addition to the SM, the experimentally
observed angular distribution can be written as,
\begin{equation*}
\frac{d\Gamma}{d\cos\theta} = \Gamma^{\text{SM}} \left(\frac{3}{4} \sin^2\theta + \frac{1}{2} \epsilon_0 \right),
\end{equation*}
where $\epsilon_0 = \Gamma^{\text{NP}}/\Gamma^{\text{SM}}$ is the new physics
parameter which can vary in the range $\left[0,1\right]$ if we assume the NP
contribution to be as large as that from the SM. Doing integration over
$\cos\theta$ we get, $\Gamma = \Gamma^{\text{SM}} \left(1+\epsilon_0\right) =
\Gamma^{\text{SM}} + \Gamma^{\text{NP}}$. This implies
\begin{equation}\label{eq:Scalar-NP-Dist-B2Knn}
\frac{1}{\Gamma} \frac{d\Gamma}{d\cos\theta} = \frac{3\sin^2\theta + 2
	\epsilon_0}{4 \left(1+\epsilon_0\right)}.
\end{equation}
Since Eq.~\eqref{eq:Scalar-NP-Dist-B2Knn} is identical to
Eq.~\eqref{eq:Scalar-NP-Dist-B2Plnu}, the angular distribution for this case is
also as shown in Fig.~\ref{fig:Scalar-NP-B2Plnu} where we have varied
$\epsilon_0$ in the range $[0,1]$. Once again at two specific values of
$\cos\theta$, namely $\cos\theta = \pm 1/\sqrt{3} \approx \pm 0.57735$
corresponding to the angle $\theta \approx 54.74^{\circ}$, there is no
difference between the standard model prediction alone and the combination of
standard model and scalar new physics contribution. At these two points in
$\cos\theta$, the normalized uni-angular distribution always has the value
$0.5$, even if there is some scalar new physics contributing to our process
under consideration.

Since the angular distribution as shown in Eq.~\eqref{eq:Scalar-NP-Dist-B2Knn}
is fully symmetric under $\cos\theta \leftrightarrow -\cos\theta$, the number of
events in the four segments of the Dalitz plot (equivalently in the four
$\cos\theta$ bins) satisfy the following relationship,
\begin{equation}\label{eq:Scalar-NP-bins}
\frac{N_I}{N_{II}} = \frac{5+8\epsilon_0}{11 + 8\epsilon_0} =
\frac{N_{IV}}{N_{III}}.
\end{equation}
It is easy to see that $\epsilon_0=0$ gives the SM prediction of
Eq.~\eqref{eq:SM-bins} as expected.

\paragraph{\textbf{Vector type new physics:}}

Let us now discuss another new physics scenario, such as the case of a
flavor-changing $Z'$ or a dark photon $\gamma_D$ giving rise to the final pair
of fermions $f_1 f_2$. We assume that for this kind of new physics scenario,
$F_V^+ = F_V^{\text{NP}} \neq 0$ and other form factors are zero. For this kind
of new physics we get,
\begin{align*}
C_0^{\text{NP}} =& 2 \modulus{F_V^{\text{NP}}}^2
\lambda\left(s,m_B^2,m_K^2\right),\\%
C_1^{\text{NP}} =& 0,\\%
C_2^{\text{NP}} =& -8 b^2 \modulus{F_V^{\text{NP}}}^2,
\end{align*}
where $b = \left(\sqrt{\left(s-4m^2\right)} \,
\sqrt{\lambda\left(s,m_B^2,m_K^2\right)}\right)/\left(2\sqrt{s}\right)$ and
$4m^2 \leqslant s \leqslant \left(m_B-m_K\right)^2$. The angular distribution
for the NP alone contribution can, therefore, be written in terms of
$T_0^{\textrm{NP}}$ and $T_2^{\textrm{NP}}$ which are directly proportional to
$C_0^{\text{NP}}$ and $C_2^{\text{NP}}$ respectively. It would lead us to
describe the complete angular distribution in terms of $T_0^{\text{NP}}$ and
$\epsilon=\Gamma^{\textrm{NP}}/\Gamma^{SM}$ using
Eq.~\eqref{eq:Tensor-NP-Dist-B2Plnu} and the angular distribution would look
like the one shown in Fig.~\ref{fig:Tensor-NP}. However, it is possible to
describe the effects of NP on the angular distribution using a different set of
parameters as well. For this we start a fresh with the angular distribution for
the NP contribution alone, which in our case is given by
\begin{equation*}
\frac{d^2\Gamma^{\text{NP}}}{ds \, d\cos\theta} = \frac{b \modulus{F_V^{\text{NP}}}^2 \lambda\left(s,m_B^2,m_K^2\right) \, \left( s \sin^2\theta + 4m^2 \cos^2\theta \right)}{64 \, \pi^3 \, m_B^2 \left(m_B^2 - m_K^2
	+ s \right) \sqrt{s}}.
\end{equation*}
Doing integration over $\cos\theta$ we obtain,
\begin{equation*}
\frac{d\Gamma^{\text{NP}}}{ds} = \frac{b \modulus{F_V^{\text{NP}}}^2
	\lambda\left(s,m_B^2,m_K^2\right)}{64 \, \pi^3 \, m_B^2 \left(m_B^2 - m_K^2 + s
	\right) \sqrt{s}} \left( \frac{4s + 8m^2}{3} \right).
\end{equation*}
Therefore, the normalized uni-angular distribution is given by
\begin{equation}\label{eq:Vector-NP-Dist-s-B2Knn}
\frac{1}{d\Gamma^{\text{NP}}/ds} \frac{d^2\Gamma^{\text{NP}}}{ds \, d\cos\theta} = \frac{3}{4} \left(\frac{s \sin^2\theta + 4m^2 \cos^2\theta}{s + 2m^2}\right).
\end{equation}
It is interesting to compare this with the standard model expression,
\begin{equation}\label{eq:SM-Dist-s-B2Knn}
\frac{1}{d\Gamma^{\text{SM}}/ds} \frac{d^2\Gamma^{\text{SM}}}{ds \, d\cos\theta} = \frac{3}{4} \sin^2\theta.
\end{equation}

\begin{figure*}[hbtp]
\centering%
\includegraphics[scale=0.8]{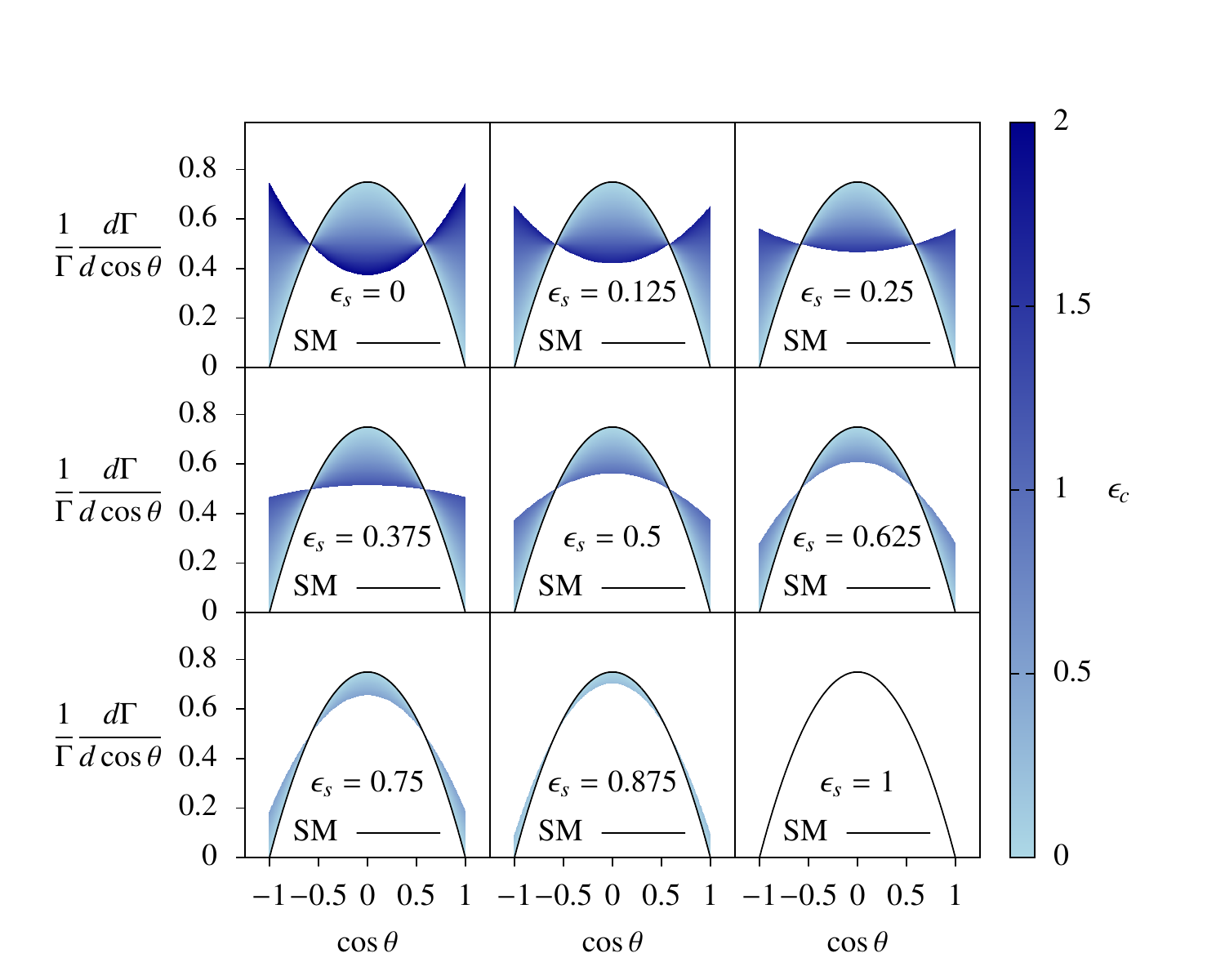}%
\caption{Normalized uni-angular distribution showing the effect of a vector new
	physics contribution to $B \to K f_1^{\CheckMark} f_2^{\CrossMark}$. }%
\label{fig:Vector-NP}
\end{figure*}

Since the range for $s$ is different in the SM and the NP scenarios, we can not
add Eqs.~\eqref{eq:Vector-NP-Dist-s-B2Knn} and \eqref{eq:SM-Dist-s-B2Knn}
directly. Carrying out the integration over $s$ we get,
\begin{equation*}
\frac{d\Gamma^{\text{NP}}}{d\cos\theta} = \frac{3}{4} \Big( \mathcal{S}
\sin^2\theta + \mathcal{C} \cos^2\theta \Big),
\end{equation*}
where
\begin{align*}
\mathcal{S} &= \int_{4m^2}^{(m_B-m_K)^2} \frac{d\Gamma^{\text{NP}}}{ds}
\left(\frac{s}{s+2m^2}\right) ds,\\%
\mathcal{C} &= \int_{4m^2}^{(m_B-m_K)^2} \frac{d\Gamma^{\text{NP}}}{ds}
\left(\frac{4m^2}{s+2m^2}\right) ds.
\end{align*}
Doing integration over $\cos\theta$ we get,
\begin{equation*}
\Gamma^{\text{NP}} = \mathcal{S} + \mathcal{C}/2,
\end{equation*}
and hence
\begin{equation*}
\frac{1}{\Gamma^{\text{NP}}} \frac{d\Gamma^{\text{NP}}}{d\cos\theta} = \frac{3
	\left(\mathcal{S} \sin^2\theta + \mathcal{C} \cos^2\theta\right)}{2
	(2\mathcal{S} + \mathcal{C})}.
\end{equation*}
For the SM contribution we know that
\begin{equation*}
\frac{1}{\Gamma^{\text{SM}}} \frac{d\Gamma^{\text{SM}}}{d\cos\theta} = \frac{3}{4} \sin^2\theta.
\end{equation*}
Now the uni-angular distribution for the process $B \to K f_1^{\CheckMark}
f_2^{\CrossMark}$ is given by,
\begin{equation*}
\frac{d\Gamma}{d\cos\theta} =\frac{3}{4} \Gamma^{\text{SM}} \left( \left(1 +\epsilon_s \right) \sin^2\theta + \epsilon_c \cos^2\theta \right),
\end{equation*}
where $\epsilon_s = \mathcal{S}/\Gamma^{\text{SM}}$ and $\epsilon_c =
\mathcal{C}/\Gamma^{\text{SM}}$, are the two parameters which describe the
effect of vector type NP. It is easy to check that,
\begin{equation*}
\Gamma = \frac{3}{4} \Gamma^{\text{SM}} \left( \frac{4}{3} \left(1+\epsilon_s\right) + \frac{2\epsilon_c}{3} \right) = \Gamma^{\text{SM}} + \Gamma^{\text{NP}}.
\end{equation*}
Therefore, the normalized angular distribution is given by,
\begin{equation}\label{eq:Vector-NP-Dist-B2Knn}
\frac{1}{\Gamma} \frac{d\Gamma}{d\cos\theta} = \frac{3 \left(1 + \epsilon_s\right) \sin^2\theta + 3\epsilon_c \cos^2\theta}{4 \left(1+\epsilon_s\right) + 2 \epsilon_c}.
\end{equation}
It is important to note that, if we consider the mass of the fermion $f$ to be
zero, i.e.\ $m=0$, then $\epsilon_c = 0$, since $\mathcal{C} =0$. In such a case
the uni-angular distribution is given by,
\begin{equation*}
\frac{1}{\Gamma} \frac{d\Gamma}{d\cos\theta} = \frac{3}{4} \sin^2\theta, \qquad \left(\text{here } \epsilon_c=0\right)
\end{equation*}
which is same as that of the SM case. This is plausible, as in the SM case also
one has $m=0$ for the neutrino mass and only vector and axial-vector currents
contribute.

Assuming that the NP contribution can be smaller than or as large as the SM
contribution, i.e.\ $0 \leqslant \Gamma^{\text{NP}} \leqslant
\Gamma^{\text{SM}}$, we get
\begin{equation*}
0 \leqslant \epsilon_s + \epsilon_c/2 \leqslant 1.
\end{equation*}
Thus $0 \leqslant \epsilon_s \leqslant 1$ implies that $0 \leqslant \epsilon_c
\leqslant 2(1-\epsilon_s)$.

In Fig.~\ref{fig:Vector-NP} we have considered nine values of $\epsilon_s$ and
varied $\epsilon_c$ in the range $[0,2\left(1-\epsilon_s\right)]$, to obtain the
uni-angular distribution. It is clearly evident in Fig.~\ref{fig:Vector-NP} that
$\epsilon_c=0$ case is always indistinguishable from the SM case, as it should
be. Just like the scalar-type new physics case, we observe that at $\cos\theta =
\pm 1/\sqrt{3} \approx \pm 0.57735$, there is no difference between the SM
prediction alone and the combination of SM and NP contributions.

It is also easy to notice that the angular distribution as given in
Eq.~\eqref{eq:Vector-NP-Dist-B2Knn} is symmetric under $\cos\theta
\leftrightarrow -\cos\theta$, and solving for the number of events in the four
segments of the Dalitz plot (equivalently the four $\cos\theta$ bins) we get,
\begin{equation}
\frac{N_{I}}{N_{II}} = \frac{5 \left(1+\epsilon_s\right) +
	7\epsilon_c}{11\left(1+\epsilon_s\right) + \epsilon_c} = \frac{N_{IV}}{N_{III}}.
\end{equation}
It is easy to see that when $\epsilon_c = 0 = \epsilon_s$ we get back the SM
prediction of Eq.~\eqref{eq:SM-bins} as expected.

\subsection{Discussion}

It should be noted that our discussions on the types of NP contributions to the
S2 and GS2 modes, specifically $B \to P \ell^- f^{\CrossMark}$ and $B \to K
f_1^{\CheckMark} f_2^{\CrossMark}$ respectively, has been fully general. There
is no complications arising out of hadronic form factors since we have
considered normalized angular distribution. It should be noted that our analysis
also does not depend on how large or small the masses of the fermions
$f,f_{1,2}$ are, as long as they are non-zero.

It is also very interesting to note that both the scalar and tensor type of NP
for the $B \to P \ell^- f^{\CrossMark}$ decays and both the scalar and vector
types of NP for the $B \to K f_1^{\CheckMark} f_2^{\CrossMark}$ decays, exhibit
similar behaviour at $\cos\theta = \pm 1/\sqrt{3}$. In order to know the real
reason behind this we must do a very general analysis. Let us assume that the
most general angular distribution for the processes $B \to P \ell^-
f^{\CrossMark}$ and $B \to K f_1^{\CheckMark} f_2^{\CrossMark}$ is given by
Eq.~\eqref{eq:Gen-ang-dist}. If we now equate this distribution to the SM
prediction of Eq.~\eqref{eq:SM-Dist-B2Plnu-massless} or
Eq.~\eqref{eq:SM-Dist-B2Knn}, and solve for $\cos\theta$ after substituting
Eq.~\eqref{eq:Def-T012} we find that,
\begin{equation}\label{eq:costheta-gen-sol}
\cos\theta = \frac{-c_1 \pm \sqrt{c_1^2 + 3
		\left(c_0+c_2\right)^2}}{3\left(c_0+c_2\right)},
\end{equation}
where the $c_j$'s (for $j=0,1,2$) are obtained from Eq.~\eqref{eq:cj} with
appropriate substitutions of masses and form factors. Thus
Eq.~\eqref{eq:costheta-gen-sol} is the most general solution that we can get for
the two specific values of $\cos\theta$. However, let us look at the specific
case when $c_1=0$. Only in this situation do we get
\begin{equation}
\cos\theta = \pm 1/\sqrt{3}.
\end{equation}
Now it is clear that since, in both the scalar and tensor type of NP
considerations for the $B \to P \ell^- f^{\CrossMark}$ decays and in both the
scalar and vector types of NP considerations for the $B \to K f_1^{\CheckMark}
f_2^{\CrossMark}$ decays, the angular distribution did not have any term
directly proportional to $\cos\theta$ (i.e.\ $c_1=0$), we obtained the same
$\cos\theta = \pm 1/\sqrt{3}$ result in both the cases. Therefore, if the
observed normalized uni-angular distribution does not have the value $0.5$ at
$\cos\theta = \pm 1/\sqrt{3}$, it implies that $c_1 \neq 0$.

Another interesting aspect of the two specific NP contributions we have
considered, is that from Figs.~\ref{fig:Scalar-NP-B2Plnu}, \ref{fig:Tensor-NP}
and \ref{fig:Vector-NP} one can clearly see that the vector and tensor types of
NP can accommodate a much larger variation in the angular distribution than the
scalar type NP. However, there is also a certain part of the angular
distribution for which both scalar and vector (or tensor) types of NP give
identical results. This happens when
\begin{equation}
\epsilon_0 = \frac{3\epsilon_c}{2\left(1 + \epsilon_s - \epsilon_c\right)} =  \frac{\epsilon \left( 3 - 4T_0^{\textrm{NP}} \right)}{1 - \epsilon \left( 2 - 4T_0^{\textrm{NP}} \right)}.
\end{equation}
In order for $\epsilon_0$ to vary in the range $[0,1]$ we find that (i) for
$0\leqslant \epsilon_s \leqslant 1$ we have $0 \leqslant \epsilon_c \leqslant
2\left(1+\epsilon_s\right)/5$ and (ii) for $0 \leqslant \epsilon \leqslant 1$ we
have $\frac{1}{2} \leqslant T_0^{\textrm{NP}} \leqslant \frac{3}{4}$. In these
specific regions, therefore, it would not be possible to clearly distinguish
whether scalar or vector or tensor type NP is contributing to our process under
consideration. Nevertheless, our approach can be used to constraint these NP
hypothesis without any hadronic uncertainties.

\section{Conclusion}\label{sec:conclusion}

We have shown that all NP contributions to three-body semi-hadronic decays of
the type $P_i \to P_f f_1 f_2$, where $P_{i(f)}$ denotes appropriate initial
(final) pseudo-scalar meson and $f_{1,2}$ are a pair of fermions, can be
codified into the most general Lagrangian which gives rise to a very general
angular distribution. The relevant NP information can be obtained by using
various angular asymmetries, provided at least one of the final pair of fermions
has some detectable signature, such as a displaced vertex, in the detector.
Depending on the detection feasibility of the final fermions we have grouped the
$P_i \to P_f f_1 f_2$ decays into three distinct categories: (i) S1 where both
$f_1$ and $f_2$ are detected, (ii) S2 where either $f_1$ or $f_2$ gets detected,
and (ii) S3 where neither $f_1$ nor $f_2$ gets detected. We consider the
possibility that with advancement in detector technology S3 decays could, in
future, be grouped under S2 category. We analyze some specific NP scenarios in
each of these categories to illustrate how NP affects the angular distribution.
Specifically we have analyzed (a) lepton-flavor violating S1 decay $B \to P
\ell^- \ell'^+$ (with $P = \pi, K, D$ and $\ell,\ell'=e,\mu,\tau$) showing
angular signatures of all generic NP possibilities, (b) S2 decays of the type $B
\to P \ell^- f$ (where $f$ is not detected in the laboratory) showing the effect
of a scalar type and a tensor type NP on the angular distribution, and finally
(c) S3 decays (more correctly generalized S2 decays) of the type $B \to K f
\bar{f}$ (where either $f$ or $\bar{f}$ gets detected in an advanced detector)
showing the effects of a scalar type and a vector type NP on the angular
distribution. The effects on the angular distribution can be easily estimated
from Dalitz plot asymmetries. The signatures of NP in angular distribution are
distinct once the process is chosen carefully. Moreover, as shown in our
examples it can be possible to do the identification and quantification of NP
effects without worrying about hadronic uncertainties. We are optimistic that
our methodology can be put to use in LHCb, Belle II in the study of appropriate
$B$ meson decays furthering our search for NP.

\acknowledgments 

This work was supported in part by the National Research Foundation of Korea
(NRF) grant funded by the Korean government (MSIP) (No.2016R1A2B2016112) and
(NRF-2018R1A4A1025334). This work of D.S. was also supported (in part) by the
Yonsei University Research Fund (Post Doc. Researcher Supporting Program) of
2018 (project no.: 2018-12-0145).

\end{document}